\documentclass[sn-mathphys, pdflatex]{sn-jnl}

\usepackage{bm}
\usepackage{xcolor}

\newcommand{\red}[1]{{{\color{black}#1}}}
\newcommand{\black}[1]{{{\color{black}#1}}}

\begin{document}

\title[Path large deviations of weak turbulence]{Path large deviations for the kinetic theory of weak turbulence}

\author[1]{Jules Guioth}\email{jules.guioth@ens-lyon.fr}
\author*[1]{Freddy Bouchet}\email{freddy.bouchet@cnrs.fr}
\author[2]{Gregory L. Eyink}\email{eyink@jhu.edu}

\affil*[1]{ENSL, CNRS, Laboratoire de Physique, F-69342 Lyon, France}
\affil[2]{Department of Applied Mathematics and Statistics and, Department of Physics \& Astronomy, The Johns Hopkins University, Baltimore, MD, USA}

\abstract{
We consider a generic Hamiltonian system of nonlinear interacting
waves with 3-wave interactions. In the kinetic regime of wave
turbulence, which assumes weak nonlinearity and large system size,
the relevant observable associated with the wave amplitude is the
empirical spectral density that appears as the natural precursor of
the spectral density, or spectrum, for finite system size. Following
classical derivations of the Peierls equation for the moment generating
function of the wave amplitudes in the kinetic regime, we propose
a large deviation estimate for the dynamics of the empirical spectral
density, where the number of admissible wavenumbers, which is proportional
to the volume of the system, appears as the natural large deviation
parameter. The large deviation stochastic Hamiltonian that quantifies
the minus of the log-probability of a trajectory is computed within the
kinetic regime which assumes the \emph{Random Phase} approximation
for weak nonlinearity. We compare this Hamiltonian with the one for
a system of modes interacting in a mean-field way with the empirical
spectrum. Its relationship with the \emph{Random Phase }and \emph{Amplitude}
approximation is discussed. Moreover, \black{for the specific case when no forces and dissipation are present,} a few fundamental properties of
the large deviation dynamics are investigated. We show that the latter
conserves total energy and momentum, as expected for a
3-wave interacting systems. In addition, we compute the equilibrium quasipotential
and check that global detailed balance is satisfied at the large deviation
level. \black{Finally, we discuss briefly some physical applications of the theory.}
}

\keywords{Large deviations, Weak turbulence, Wave kinetics}

\maketitle

\section{Introduction}

Waves are present in many different contexts ranging from quantum
mechanics, geophysics to general relativity. In many contexts, the
media are dispersive and waves weakly interact. When the number of
interacting waves is large, the dynamics may be of weakly interacting
phase-incoherent waves, whose amplitudes are fluctuating. The theory
of weak turbulence is a statistical theory that aims at describing
this state of weakly interacting random waves \cite{Zakharov1992,Nazarenko2011},
in a weak nonlinear regime. This kinetic theory requires two hypothesis.
First a weak nonlinear interaction, characterized by a separation
of time scales between the linear wave motion and the slow nonlinear
evolution of the amplitude. The ratio of these two time scales is
denoted $\epsilon^{2}$, and we denote the kinetic time $\tau=\epsilon^2 t$
where $t$ is the physical time. Second, a large system size, \black{ $L\gg v_g t $,
much greater than the distance travelled by waves propagating with their group
velocity $v_g=\omega'(k).$ This condition guarantees that} interactions between
waves occur through broad resonances, i.e. many modes interact close to the exact
resonant conditions, locally in Fourier space.

If $\left\lvert A_{\bm{k}} \right\rvert$ is the amplitude of waves with wavenumber
$\bm{k}$, in this kinetic limit, it is convenient to consider
the rescaled amplitude $\left\lvert a_{\bm{k}}\right\rvert=\left(\tfrac{2\pi}{L}\right)^{-d/2}\epsilon^{-1}\left\lvert A_{\bm{k}}\right\rvert$.
The theory mainly focus on the rescaled empirical spectral density
$\hat{n}(\bm{\xi},\tau)=\left(\frac{2\pi}{L}\right)^{d}\sum_{\bm{k}}\vert a_{\bm{k}}\vert ^{2}\delta(\bm{\xi}-\bm{k})$, $\bm{\xi} \in \mathbb{R}^d$,
that measures \black{in the space of distribution} how the square of the amplitude $\left\lvert a_{\bm{k}}\right\rvert^{2}$
depends on the continuous wavenumber $\bm{\xi}$. The main prediction of this
theory is the kinetic equation that describe the averaged or most
probable evolution of the empirical spectral density. In the kinetic
limit, it is expected that the empirical spectral density converges,
as a law of large numbers $\lim_{L\underset{\text{Kin}}{\rightarrow}\infty}\hat{n}(\bm{\xi},\tau)=n(\bm{\xi},\tau)$,
to a spectral density $n$ that satisfies the so called kinetic equation
(see equation (\ref{eq:kinetic_eq}) below). \black{This equation was first
derived by Peierls in the context of phonons in a crystal \cite{peierls1929kinetischen} and by
Hasselmann for classical wave systems \cite{hasselmann1962non}.}
The symbol $L\underset{\text{Kin}}{\rightarrow}\infty$
means that we take the limit $L\rightarrow\infty$ with some prescribed
conditions on $\epsilon$, to be defined later. We call it the kinetic
limit.

Already since Peierls' work, the theory has been developed to predict
fluctuations of the wave amplitude, by describing higher-order statistics
than the average. In the kinetic limit, the moment generating function
of the spectrum satisfies the Peierls equation (see \cite{Nazarenko2011}).
The paper \cite{EyinkShi2012} clarified the derivation of the evolution
of the moment generating function, by computing consistently all terms
in power of $1/L$, with a scaling that properly gives access to the
law of large numbers.  The aim of this note is to explain how those classical
results connect to large deviation theory. More precisely, we aim at
justifying that, starting from the initial condition $\hat{n}(\tau=0)=n_{0}$,
the probability that the empirical spectrum paths $\left\{ \hat{n}(\tau)\right\} _{0\leq\tau\leq \tau_{\rm fin}}$
remain close to a prescribed spectral density $\left\{ n(\tau)\right\} _{0\leq\tau\leq \tau_{\rm fin}}$
satisfies a large deviation principle

\[
\mathbb{P}_{n_{0}}\left[\left\{ \hat{n}(\tau)\right\} _{0\leq \tau\leq \tau_{\mathrm{fin}}}=\left\{ n(\tau)\right\} _{0\leq \tau \leq \tau_{\mathrm{fin}}}\right]\underset{L\underset{\text{Kin}}{\rightarrow}\infty}{\asymp}{\rm e}^{-\left(\frac{L}{2\pi}\right)^{d}\sup_{\lambda}\int_{0}^{\tau_{\mathrm{fin}}}{\rm d}\tau\,\left[\int\text{d}k\,\lambda\dot{n}-H[n,\lambda]\right]},
\]
where the large deviation Hamiltonian $H$ characterizes all fluctuation
statistics. Here we have assumed that $n(\tau=0)=n_{0}$. \black{One of the main results of the paper is to
give an explicit expression for $H$ and to study its properties}.

We formally derive in this paper this large deviation estimate for the stochastic
dynamics of the spectral density within the kinetic regime. Following
classical derivations of the Peierls equation \cite{Nazarenko2011,EyinkShi2012},
we propose an expression for the stochastic Hamiltonian. The Hamiltonian
is obtained in the kinetic limit where the phases of the modes are
assumed to be independent and uniformly distributed (\emph{Random
Phase }(RP)\emph{ }approximation). Following closely previous derivations
\cite{Nazarenko2011,EyinkShi2012} for different probabilistic quantities,
our derivation involves a formal exact computation truncated at order
2 in $\epsilon$. As we will explain in more detail, a rigorous derivation
would require the control of higher order terms. \red{Following pioneering work of Lukkarinen \& Spohn \cite{lukkarinen2011weakly,lukkarinen2009not},
recent breakthroughs in mathematical derivation of the wave kinetic equations \cite{buckmaster2021onset, deng2021full} as well as of higher-order statistics \cite{deng2021propagation} have been achieved. We believe that rigorous proof of our large-deviations theory thus may be within the realm of possibility.}

\black{In three recently published papers, dynamical large deviation principles related to the main classical kinetic theories have been established, starting from Hamiltonian dynamics. The first one \cite{Bouchet2020} dealt with the large deviations for dilute gases (associated with the Boltzmann equation), the second one dealt with large deviations for plasma fluctuations at scales much \black{larger} then the Debye length \cite{feliachi2021dynamical} (associated with the Landau equation), and the third one with large deviations for particles with mean field interactions \cite{feliachi2022dynamical} (associated with the Balescu--Guernsey--Lenard equation). The present paper gives a similar result for the kinetic theory of wave turbulence. Those four results describe large deviation principles which are analogous to the macroscopic fluctuation theory for diffusive systems \cite{eyink1990dissipation,bertini2015macroscopic}. The main difference is that the derivation starts from the Hamiltonian reversible dynamics, rather than from Markov stochastic processes for particles. The work \cite{Bouchet2020} also introduces the general formalism of path large deviation properties, and explains all the properties of the large deviation Hamiltonian that should be expected for any kinetic theory, in relation with conservation laws, the increase of entropy, the irreversibility paradox, and detailed balance related to time reversal symmetry.} \\

The paper is organized as follows. We briefly present in section \ref{sec:Hamiltonian_dyn}
the Hamiltonian dynamics of the waves, in a Fourier decomposition.
Section \ref{sec:large_deviations_spectral_density} is the core of
the paper. After introducing the empirical spectral density and the
joint limit of weak nonlinearity and large system size that form the
kinetic limit, we derive the large deviation Hamiltonian for the empirical
spectrum. We then discuss
the conservation laws as well as the equilibrium quasipotential and
the time-reversal symmetry at the large deviations level in Section
\ref{sec:hamiltonian_properties}.
Inspired by the \emph{Random Phase }and \emph{Amplitude}
(RPA) approximation, it might be natural to consider the large deviations
Hamiltonian for a system of modes that \black{evolve via a mean-field Langevin
model with interactions only} through the empirical spectrum. \black{This mean-field
approach was shown previously in \cite{EyinkShi2012} (section 3.2.2) to yield
the exact results for probability distribution of single-mode amplitudes
\cite{Choi_2005,Nazarenko2011,deng2021propagation}. However,
we explain that for large deviations} the actual wave-turbulence
Hamiltonian and this mean-field Hamiltonian are different. A short
comparison between both Hamiltonians \red{as well as the fluctuations is provided}.
\red{Finally, we briefly discuss in section \ref{sec:out-of-eq} how one can extend the path large deviation theory of Sec. \ref{sec:large_deviations_spectral_density} to out-of-equilibrium situations. A short note on (typical) Gaussian fluctuations is also provided.}
\black{The section \ref{sec:Conclusion}
summarizes our results and discusses potential applications to noise-induced
transitions and spontaneous symmetry-breaking in wave turbulence.}
 For the sake of completeness, some details of the relevant
computations of the paper are briefly presented in Appendices \ref{sec:Appendix-B}
and \ref{sec:Appendix-Quasipotential_and_Entropy}.

\section{Hamiltonian dynamics of nonlinear waves}\label{sec:Hamiltonian_dyn}

We follow the classical literature, for instance \cite{Nazarenko2011}
and references therein, to describe the Hamiltonian dynamics of waves.
For simplicity, we assume that the waves are described by a scalar
field $\Psi(\bm{x},t)$, where $\bm{x}$ is the position in a space
of dimension $d$ and $t$ is time. We will consider the simplest
case where the signal $\Psi$ satisfies a dispersive wave equation
with a quadratic nonlinear (or interaction) term. Exchange of energy
between modes will then mainly occur through 3-wave interactions.
The results of this discussion would simply generalize to other cases.

\subsection{Fourier modes decomposition}

For simplicity, the field $\Psi(\bm{x},t)$ is assumed to be periodic
in all $d$ directions with a period $L$. We note $\mathbb{V}_{L}^{d}=[0,L]^{d}$
a representative volume of the system. Following \cite{Nazarenko2011,EyinkShi2012}
we will adopt the classic Hamiltonian formalism in the Fourier space.
We define the Fourier series decomposition of $\Psi(\bm{x},t)$ as
\begin{align}
A_{\bm{k}}(t) & =\frac{1}{L^{d}}\int_{\mathbb{V}_{L}^{d}}\Psi(\bm{x},t){\mathrm{e}^{-i\bm{k}\cdot\bm{x}}}{\rm d}^d \bm{x}\label{eq:fourier_decomposition} \qquad \left(\bm{k} \in \mathbb{Z}_L^d  \right)\\
\Psi(\bm{x},t) & =\sum_{\bm{k} \in \mathbb{Z}_L^d}A_{\bm{k}}(t){\mathrm{e}^{i\bm{k}\cdot\bm{x}}}\qquad \left(\bm{x} \in \mathbb{V}_L^d  \right) \nonumber
\end{align}
with $\mathbb{Z}_L^d= \tfrac{2\pi}{L}\mathbb{Z}^d$. Unless stated otherwise,
we will adopt the notation $\sum_{\bm{k}}\equiv\sum_{\bm{k}\in \mathbb{Z}_L^d }$.

\paragraph{}

For some specific applications or discussions, it might be important
to discuss small scale regularization. For instance, in equilibrium,
as briefly detailed in Appendix \ref{sec:Appendix-Quasipotential_and_Entropy},
it is known since the Rayleigh-Jeans paradox for black body radiation
that a system of waves in the absence of dissipation at high wavenumber
generally leads to ultraviolet divergences \cite{Nazarenko2011,Zakharov1992}.
Then one might need to introduce a maximum wavenumber $k_{{\rm max}}$
(sharp cut-off) such that $A_{\bm{k}}=0$ for $\left\vert  \bm{k}\right\vert  >k_{{\rm max}}$.
Since $k_{{\rm min}}=\tfrac{2\pi}{L}$, one thus gets a finite number
of modes $\mathcal{N}_{L}\approx\left(k_{{\rm max}}/k_{{\rm min}}\right)^{d}\propto\left(\tfrac{L}{2\pi}\right)^{d}$.
However, for most cases of interest, for instance for fluids when small
scale dissipation is present, this assumption is not required for
the study of the dynamics, as long as the sum over $\bm{k}\in \mathbb{Z}_L^d$
converges. In the following, unless otherwise stated we will consider
sums over the full space $\mathbb{Z}_L^d$ in the sequel,
assuming convergence of these sums.

\subsection{Hamiltonian dynamics }

Following the notations of \cite{EyinkShi2012}, we consider a generic
Hamiltonian restricted to 3-wave interaction

\begin{equation}
\mathcal{H}=\sum_{\bm{k}}\omega_{\bm{k}}A_{\bm{k}}A_{\bm{k}}^{\ast}\;+\sum_{\sigma_{1},\sigma_{2},\sigma_{3}}\sum_{\bm{k}_{1},\bm{k}_{2},\bm{k}_{3}}V_{\bm{k}_{1}\bm{k}_{2}\bm{k}_{3}}^{\sigma_{1}\sigma_{2}\sigma_{3}}A_{\bm{k}_{1}}^{\sigma_{1}}A_{\bm{k}_{2}}^{\sigma_{2}}A_{\bm{k}_{3}}^{\sigma_{3}}\delta_{\sigma_{1}\bm{k}_{1}+\sigma_{2}\bm{k}_{2}+\sigma_{3}\bm{k}_{3},0}\label{eq:H_dynamics}
\end{equation}
where $\sigma_{i}=\pm1$, $A_{\bm{k}}^{+}=A_{\bm{k}}$ and $A_{\bm{k}}^{-}=A_{\bm{k}}^{\ast}$,
with $A^{\ast}$ being the complex conjugate of $A$. We will use
the following abbreviated notation $\vec{\sigma}=(\sigma_{1,}\sigma_{2},\sigma_{3})$,
$\vec{\bm{k}}=(\bm{k}_{1,}\bm{k}_{2},\bm{k}_{3})$ and $\vec{\sigma}\cdot\vec{\bm{k}}=\sigma_{1}\bm{k}_{1}+\sigma_{2}\bm{k}_{2}+\sigma_{3}\bm{k}_{3}$.
Also, we denote $\mathcal{H}_{2}=\sum_{\bm{k}}\omega_{\bm{k}}A_{\bm{k}}A_{\bm{k}}^{\ast}$
the quadratic part of the Hamiltonian, and $\mathcal{H}_{3}=\sum_{\vec{\sigma}}\sum_{\vec{\bm{k}}}V_{\vec{\bm{k}}}^{\vec{\sigma}}A_{\bm{k}_{1}}^{\sigma_{1}}A_{\bm{k}_{2}}^{\sigma_{2}}A_{\bm{k}_{3}}^{\sigma_{3}}\delta_{\vec{\sigma}\cdot\vec{\bm{k}},0}$
the cubic part. We note that the Hamiltonian $\mathcal{H}$ is an
energy density (an energy per unit of volume). We will assume that
it remains finite when $L\to\infty$.

We assume the two following properties

\begin{align}
\left(V_{\vec{\bm{k}}}^{\vec{\sigma}}\right)^{\ast} & =V_{\vec{\bm{k}}}^{-\vec{\sigma}}\qquad\text{(}\text{reality of the Hamiltonian}) \label{eq:prop_V_hamiltonian} \\
V_{\pi(\vec{\bm{k}})}^{\pi(\vec{\sigma})} & =V_{\vec{\bm{k}}}^{\vec{\sigma}}\qquad\text{(permutation symmetry)} \nonumber
\end{align}
for any permutation $\pi$ of the triplets $(1,2,3)$ (\emph{i.e.
}$\pi(\vec{\sigma})=(\sigma_{\pi(1)},\sigma_{\pi(2)},\sigma_{\pi(3)})$
and $\pi(\vec{\bm{k}})=(\bm{k}_{\pi(1)},\bm{k}_{\pi(2)},\bm{k}_{\pi(3)})$).
The first assumption ensures that the Hamiltonian is a real number.
The second can always be assumed without loss of generality.

The evolution equation (Hamilton equations) in the Fourier space reads
as
\begin{align}
i\frac{\text{d}A_{\bm{k}_{1}}}{\text{d}t} & =\frac{\partial\mathcal{H}}{\partial A_{\bm{k}_{1}}^{\ast}}=\omega_{\bm{k}_{1}}A_{\bm{k}_{1}}+3\sum_{\substack{\vec{\sigma}\\
\sigma_{1}=-1
}
}\sum_{\bm{k}_{2},\bm{k}_{3}}V_{\vec{\bm{k}}}^{\vec{\sigma}}A_{\bm{k}_{2}}^{\sigma_{2}}A_{\bm{k}_{3}}^{\sigma_{3}}\delta_{\vec{\sigma}\cdot\vec{\bm{k}},0}\quad.\label{eq:Ak_evol_eq}
\end{align}

\subsection{Dynamics for weak wave turbulence}\label{subsec:dynamics_wave_turbulence}

The weak turbulence theory is a perturbative nonlinear expansion which
is valid for small amplitudes of the field $\left\{ A_{\bm{k}}\right\} $.
This small amplitude assumption is actually a condition on the time
scale decoupling between the dynamics of the phases and the amplitudes.
We will discuss this in more detail in Sec. \ref{subsec:The-kinetic-limit}.

Actually, the reason for the amplitude of the field $\left\{ A_{\bm{k}}\right\} $
to be small is twofold. First, we assume that for any fixed $\epsilon$,
the\textbf{ }total energy density $\mathcal{H}$ has a finite limit
in the limit $L\rightarrow\text{\ensuremath{\infty}}$.\textbf{ }Looking
at the Fourier decomposition (\ref{eq:fourier_decomposition}), we
see that this is the case\footnote{Considering for instance the quadratic term $\mathcal{H}_{2}=\sum_{\bm{k}}\omega_{\bm{k}}\left\lvert A_{\bm{k}}\right\rvert^{2},$
one sees that $\left\lvert A_{\bm{k}}\right\rvert^{2}$ must be of order $L^{-d}$
so that the sum converges in the limit $L\to\infty$} when the square amplitudes $ \left\lvert A_{\bm{k}}\right\rvert^{2} $
scale as $\left\lvert A_{\bm{k}}\right\rvert^{2}\underset{L\to\infty}{\sim}\left(\tfrac{2\pi}{L}\right)^{d}$.
On top of this large-$L$ scaling we want to ensure the time scale
decoupling between the phase dynamics and the amplitudes one \cite{Nazarenko2011}.
To do so, we introduce an extra parameter $\epsilon$ $(\epsilon\ll1)$,
\emph{a priori} independent from $L$. We thus define

\begin{equation}
a_{\bm{k}}(t)=\left(\frac{L}{2\pi}\right)^{d/2}\epsilon^{-1}A_{\bm{k}}(t).\label{eq:Fourier_modes_rescaling}
\end{equation}
where $a_{\bm{k}}$ will be of order one in the limits $L\rightarrow\infty$
and $\epsilon\rightarrow0$.

Performing this change of variable in the equation of motion (\ref{eq:Ak_evol_eq})
yields

\begin{align}
i\frac{\text{d}a_{\bm{k}_{1}}}{\text{d}t} & =\omega_{\bm{k}_{1}}a_{\bm{k}_{1}}+3\epsilon\left(\frac{2\pi}{L}\right)^{d/2}\sum_{\substack{\vec{\sigma}\\
\sigma_{1}=-1
}
}\sum_{\bm{k}_{2},\bm{k}_{3}}V_{\vec{\bm{k}}}^{\vec{\sigma}}a_{\bm{k}_{2}}^{\sigma_{2}}a_{\bm{k}_{3}}^{\sigma_{3}}\delta_{\vec{\sigma}\cdot\vec{\bm{k}},0}\quad.\label{eq:ak_evol_eq}
\end{align}
It is useful to use the interaction representation. We thus perform
the change of variables $b_{\bm{k}}(t)=a_{\bm{k}}(t)e^{i\omega_{\bm{k}}t}$
and obtain

\begin{align}
i\frac{\text{d}b_{\bm{k}_{1}}}{\text{d}t} & =3\epsilon\left(\frac{2\pi}{L}\right)^{d/2}\sum_{\substack{\vec{\sigma}\\
\sigma_{1}=-1
}
}\sum_{\bm{k}_{2},\bm{k}_{3}}V_{\vec{\bm{k}}}^{\vec{\sigma}}b_{\bm{k}_{2}}^{\sigma_{2}}b_{\bm{k}_{3}}^{\sigma_{3}}{\rm e}^{-i\left(\vec{\sigma}\cdot\vec{\omega}\right)t}\delta_{\vec{\sigma}\cdot\vec{\bm{k}},0}\:,\label{eq:bk_evol_eq}
\end{align}
with $\vec{\omega}=(\omega_{\bm{k}_{1}},\omega_{\bm{k}_{2}},\omega_{\bm{k}_{3}})$.

\section{Dynamical large deviations for the empirical spectral density in
the kinetic regime\label{sec:large_deviations_spectral_density}}

In this section, we compute the dynamical large deviations for the
empirical spectral density in the kinetic limit. The basic object
of our theory is the empirical spectral density $\hat{n}$, defined
in section \ref{subsec:def_empirical_spectral_density}. It quantifies
the mode amplitudes $\left\lvert a_{\bm{k}}\right\rvert^{2}=\left\lvert b_{\bm{k}}\right\rvert^{2}$
and how they depend on $k$. The aim of the theory is to quantify the
probability for path evolution for the spectral density. In the kinetic
limit, the fluctuations of the empirical spectral density around the
deterministic dynamics (law of large number) are small and can be
captured by a large deviation analysis. As we will see, the large
deviation speed will be $\left(L/2\pi\right)^{d}$. In section \ref{subsec:The-kinetic-limit}
we discuss more precisely the conditions when $\epsilon\to0$ and
$L\to\infty$ which define the kinetic regime. The fundamental object
for the large deviations dynamics is the stochastic Hamiltonian $H$
(defined in Eq. \eqref{eq:def_Hamiltonian} below). It corresponds to
the scaled cumulant generating function of the elementary time increment
of the empirical spectral density. It is computed in section \ref{subsec:derivation_large_deviations_Hamiltonian}.

\subsection{Definition of the empirical spectral density}\label{subsec:def_empirical_spectral_density}

Noting that the mode wavenumbers $\bm{k}\in \mathbb{Z}_L^d$
change with $L$, and are getting closer in the large $L$ limit to
form a continuum, it is natural to define the empirical spectral density

\begin{equation}
\hat{n}(\bm{\xi},t)=\left(\frac{2\pi}{L}\right)^{d}\sum_{\bm{k}}\left\lvert a_{\bm{k}}(t)\right\rvert^{2}\delta( \bm{\xi}-\bm{k} ),\label{eq:def_empirical_spectrum}
\end{equation}
where $\delta$ is a Dirac distribution in the $d$-dimensional space
of wavenumbers. The spectrum $\hat{n}$ is a distribution in the $d$-dimensional
space of wavenumbers. It is normalized such that $\epsilon^{2}$ multiplied
by its integral is the density of the integral of the square of the
field $\psi$:
\[
\int_{\mathbb{R}^d} \!\!\! \hat{n} =\frac{1}{\epsilon^{2}L^{d}}\int_{\mathbb{V}_{L}^{d}}\left\lvert \Psi\right\rvert^{2}.
\]
With this definition, we expect $\hat{n}(\bm{\xi})$ to satisfy a law
of large number in the limit $L\rightarrow\infty$ (deterministic
or continuous limit): $\hat{n}(\bm{\xi}) \underset{\mathrm{Kin.}}{\rightarrow} n\left(\bm{\xi}\right)$.
Our main goal in this paper is to quantify the fluctuations (of order
$1$ in $L$) of the empirical density $\hat{n}$, around the law of
large number.

\subsection{The kinetic limit for weak wave turbulence}\label{subsec:The-kinetic-limit}

We now detail the kinetic regime within which the law of large number
(kinetic equation) and the large deviations of the empirical spectral
density will be computed.

The kinetic regime actually requires two conditions on the elementary
time increment $\Delta t$. First, the \emph{Random Phase} (RP) approximation
\black{necessitates} the limit $\epsilon\to0$. On the other hand, the kinetic
limit \emph{per se }(valid for $\epsilon\ll1$ and large-$L$) appears
as a requirement on the number of (quasi)resonances that contribute
to the evolution of the spectrum.

We provide some more details on these conditions in the two following
subsections.

\subsubsection{Homogenization, kinetic time and the random phase approximation}\label{subsec:The-kinetic-limit}

From the weak nonlinearity assumption $\epsilon\ll1$ in equations
(\ref{eq:ak_evol_eq},\ref{eq:bk_evol_eq}), one naturally expects
a decoupling between the \emph{fast} dynamics of the phases and the
\emph{slow }dynamics of the amplitudes. This is at the basis of the kinetic
theory \cite{Nazarenko2011,Zakharov1992}. The kinetic theory can
be interpreted as a homogenization problem, where one seeks at deriving
an effective equation for the slow evolution of the mode amplitudes.
Because the phases evolving by (\ref{eq:ak_evol_eq}) are just transported at
leading order, the natural invariant measure for the phases as leading
order is a uniform measure. Then the phases are assumed to be uniformly
distributed and independent at leading order. This corresponds to
the so-called \emph{Random Phase (RP)} approximation.

\black{As has long been understood \cite{spohn2006phonon,newell2011wave}},
a full justification of the kinetic equation would require to assess
how the nonlinear dispersion relation and/or the effects of the chaotic
nonlinear dynamics, leads to the convergence to this uniform distribution
for the phases, within a very short time compared to the typical time
of the nonlinear evolution of the amplitudes
This mixing condition
would justify to forget the information about the initial condition.
It is beyond the scope
of this discussion: we will assume that the dynamics is mixing. More
precisely, we assume that for any mode with wavenumber $\bm{k}$,
mixing of the phase statistics is reached over a characteristic time
$t_{d}(\bm{k},\epsilon)$ ($t_{d}$ might depend on $\hat{n}$). Then,
for times much larger than $t_{d}(\bm{k},\epsilon)$, one can estimate
at leading order any time-integrated observable using the \emph{Random
Phase} (RP) approximation.

For small $\epsilon$, a Markov dynamics for the amplitudes can be
estimated at the dominant order in $\epsilon$ by using the RP approximation.
For systems with 3-wave interactions, one can show \cite{Nazarenko2011,Zakharov1992}
(some details are provided in Appendix \ref{sec:Appendix-B}) that
terms of order $\epsilon$ vanish in average and that the first non
trivial contribution is of order $\epsilon^{2}$ . The characteristic
time for the nonlinear evolution of the amplitudes thus appears to
scale as $1/\epsilon^{2}$ in the limit $\epsilon\to0$. More precisely,
one can introduce for each wavenumber $\bm{k}$ a nonlinear characteristic
time $t_{{\rm NL}}(\bm{k})=\tau_{{\rm NL}}(\bm{k})/\epsilon^{2}$.
The characteristic time $\tau_{{\rm NL}}(\bm{k})$ is expected to
be independent of $\epsilon$, asymptotically for small $\epsilon$.
However it is natural to expect that it depends on $\hat{n}$. It can
be estimated \emph{a posteriori }using the kinetic equation (see Eq.
\eqref{eq:kinetic_eq} henceforth). We do not provide any precise estimate
of $\tau_{{\rm NL}}(\bm{k})$ here, but some can be found in \cite[Sec. 10.2]{Nazarenko2011}.

It thus appears natural to make the change of variable $\tau=\epsilon^{2}t$
and to consider the empirical spectral density $\hat{n}$ as a function
of $\tau$ (instead of the microscopic time $t$). Finally, the condition
for the elementary time increment $\Delta t$ for the random phase
approximation to be valid, and the spectrum not to have evolved much
are: $t_{d}(\bm{k},\epsilon)\ll\Delta t\ll t_{{\rm NL}}(\bm{k})$,
for each mode $\bm{k}$. In terms of the slow time scale $\tau$,
one gets
\begin{equation}
\epsilon^{2}t_{d}(\bm{k},\epsilon)\ll\Delta\tau\ll\tau_{{\rm NL}}(\bm{k}).\label{eq:RP_condition}
\end{equation}
The mixing condition $\epsilon^{2}t_{d}(\bm{k},\epsilon)\ll\Delta\tau$
and the condition for convergence of the statistics before a nonlinear
evolution occurs $\Delta\tau\ll\tau_{{\rm NL}}(\bm{k})$ explain why
one expects a Markov dynamics for the effective dynamics of the empirical
spectrum. Equation (\ref{eq:RP_condition}) can be referred to as the
Markov condition for the kinetic regime. This is the first condition
for the kinetic regime to exist.

\subsubsection{Kinetic condition: large number of (quasi)-resonances}

As will be clear with the computation below, the effective long term
evolution of the empirical spectrum is dominated by wavenumber triads
$(\bm{k}_{1},\bm{k}_{2},\bm{k}_{3})$ that are sufficiently close
to the resonance conditions which are, respectively

\begin{equation}
\begin{cases}
\sum_{i=1}^{3}\sigma_{i}\bm{k}_{i}=0, & \text{and}\\
\sum_{i=1}^{3}\sigma_{i}\omega_{\bm{k}_{i}}=0.
\end{cases}\label{eq:resonance_cond}
\end{equation}
Whether those conditions can be met by a large number of triads $(\bm{k}_{1},\bm{k}_{2},\bm{k}_{3})$,
or just a few, or not at all, delineate different dynamical regime
(see for instance \cite{Nazarenko2011}). The so-called kinetic regime
of interest in this work is the regime for which an infinite number
of modes close to the resonant condition do contribute. This defines
a second condition for the kinetic regime beside the first condition
of mixing/time scale separation.

In the limit $L\to\infty$, at fixed $\epsilon$, the number of modes
become infinite and the resonant condition is met by an infinity of
triads. We now give a more precise range of validity for the kinetic
regime. At finite $L$, $\bm{k}\in \mathbb{Z}_L^{d}$ and the wavenumber
spacing (the distance between two close-by wavenumbers) is $2\pi/L$.
We define the frequency spacing $\Delta_{\omega}$ as the typical
distance between two discrete frequencies for the wave dynamics. For
large $L$, we have the estimate $\Delta_{\omega}\underset{L\to\infty}{\sim}\frac{2\pi}{L}\left\lvert \frac{\partial\omega}{\partial\bm{k}}\right\rvert$.
The triads that will participate to the nonlinear evolution of the
spectrum over a time window $\Delta t$ must be such that they verify
the broadened resonance condition $\left(\vec{\sigma}\cdot\vec{\omega}\right)\Delta t\ll1$
\cite{Nazarenko2011}. If we have the condition $\Delta_{\omega}\Delta t\ll1$,
we see that the broadened resonance condition $\left(\vec{\sigma}\cdot\vec{\omega}\right)\Delta t\ll1$
can be met by a large number of triads close to the exact resonant
condition $\vec{\sigma}\cdot\vec{\omega}=0$. Hence, a sufficient
condition to get a large number of resonant triads over times of order
$\Delta t=\Delta\tau/\epsilon^{2}$, where $\Delta\tau$ should be
thought as a time increment that satisfies the Markov condition (\ref{eq:RP_condition}),
is then
\begin{equation}
\frac{2\pi}{L}\left\lvert \frac{\partial\omega}{\partial\bm{k}}\right\rvert\frac{\Delta\tau}{\epsilon^{2}}\ll1.\label{eq:kinetic_condition}
\end{equation}
This is the broad resonance condition of the kinetic regime.

\subsubsection{Kinetic limit: scaling and conclusion}

Gathering the Markov (\ref{eq:RP_condition}) and broad resonance
(\ref{eq:kinetic_condition}) conditions yields

\begin{equation}
\epsilon^{2}t_{d}(\bm{k},\epsilon)\ll\Delta\tau\ll\min\left\{ \frac{L\epsilon^{2}}{2\pi}\left\lvert \frac{\partial\omega}{\partial\bm{k}}\right\rvert^{-1},\tau_{{\rm NL}}(\bm{k})\right\} .\label{eq:kinetic_ineq}
\end{equation}
\black{This condition is referred to as the kinetic regime.}

In the sequel, we will compute the statistics of the spectral density
increment $\Delta \hat{n}=\hat{n}(\cdot,\tau+\Delta\tau)-\hat{n}(\cdot,\tau)$
for an infinitesimal step $\Delta\tau$, starting from an initial
spectrum $\hat{n}(\tau=0)=n_{0}$. Hence, all computation will
be ensemble averages using the random phase distribution, conditioned
on the knowledge of $n_{0}$. Since for a given $n_{0}$ the functions
$\tau_{{\rm NL}}(\bm{k})$ can be considered as independent on $L$
and $\epsilon$ asymptotically, for a given $n_{0}$ the kinetic regime
inequalities (\ref{eq:kinetic_ineq}) can be met uniformly for small
enough $\epsilon$ and large enough $L\epsilon^{2}$ (for instance
if $L\epsilon^{2}$ is bounded from below).

Classic references \cite{Nazarenko2011,Zakharov1992} refer to the
kinetic limit as the ordered limits $L\to\infty$ and then $\epsilon\to0$.
This ensures that, asymptotically, there exists a $\Delta\tau$ such that
\eqref{eq:kinetic_ineq} is satisfied. However, we see that this condition
is too restrictive \black{compared to the condition \eqref{eq:kinetic_ineq}}
. For instance the joint limit $L\to\infty$ and
$\epsilon\to0$ such that $L\epsilon^{2}>C$, where the lower bound
$C$ is a fixed constant, is sufficient to ensure that, asymptotically,
there exists a $\Delta\tau$ such that (\ref{eq:kinetic_ineq}) is satisfied. \red{It is customary in the mathematical literature \cite{deng2021full,deng2021propagation} to introduce the scaling $\epsilon = L^{-\kappa/2}$ with $\kappa \geqslant 0$ ($\kappa = 0$ meaning that one takes the limit $L\to\infty$ before $\epsilon \to 0$). The condition $\epsilon^{2} L > C$ thus becomes $0\leqslant \kappa \leqslant 1$.}

In the following we simply consider any limit process \red{$L\to \infty$, $\epsilon \to 0$}
such that \red{$L\epsilon^{2}$ is bounded from below. We call this limit the kinetic limit, which we denote $\text{Kin.}\lim$. For any function
$\phi_{L,\epsilon}$ that depends on the parameters $L$ and $\epsilon$, we write}
\begin{equation}
  \text{Kin}\lim\phi_{L,\epsilon} = \lim_{ \substack{\epsilon \to 0,\, L\to \infty \\ \exists C>0,\, L\epsilon^2>C} } \phi_{L,\epsilon}
  \label{eq:def_kinlim}
\end{equation}

\subsection{Hamiltonian for the path large deviations}\label{subsec:derivation_large_deviations_Hamiltonian}

We now turn to the main goal of this paper that is to describe the
stochastic evolution of the empirical spectrum $\hat{n}$ in the large
deviation limit.

More precisely, we will justify that the probability (density) to
observe a trajectory for the empirical spectrum $\left\{ n(\tau)\right\} _{\tau=0}^{\tau_{{\rm f}}}$,
conditioned on the initial condition $\hat{n}(0)=n_{0}$, satisfies
a large deviation principle

\red{
\begin{equation}
\mathbb{P}_{n_{0}}\left[\left\{ \hat{n}(\tau)=n(\tau)\right\} _{0\leq\tau\leq\tau_{\rm fin}}\right]
\underset{\text{Kin}\lim}{\asymp}
{\rm e}^{-\left(\frac{L}{2\pi}\right)^{d}\mathcal{A}[n]},\label{eq:Path_Large_Deviation}
\end{equation}
}
with the action $\mathcal{A}$
\begin{equation}
\mathcal{A}[n]=\sup_{\lambda}\left\{ \int_{0}^{\tau_{{\rm f}}}{\rm d}\tau\,\left[\int \,\lambda(\bm{\xi},\tau) \dot{n}(\bm{\xi},\tau)\text{d}^d \xi - H[n,\lambda]\right]\right\} ,\;\label{eq:stochastic_action}
\end{equation}
where $\tau$ is the kinetic time $\tau=\epsilon^{2}t$ (see section
\ref{subsec:The-kinetic-limit} for a discussion of time scales).\\

For continuous-time Markov processes $\left\{ \hat{n}(\tau)\right\} _{0\leq\tau\leq \tau_{\mathrm{fin}}}$
that depend on a parameter $L$, a path large deviation principle
similar to (\ref{eq:Path_Large_Deviation}) and the expression for
$H$ can be easily obtained from the expression of the infinitesimal
generator. We now cite a general classical relation between the infinitesimal
generator of the Markov process and $H$ \cite{feng2006large,FreidlinWentzell1998},
following \cite{Bouchet2020} (section 3.1). We adapt the notations
of \cite{Bouchet2020} (section 3.1) to the specific case when $n$
is a spectral density. Therefore, \black{$n:\mathbb{R}^{d}\rightarrow\mathbb{R}$ is a distribution.}
\black{Denoting} $F$ the set of such distributions, the infinitesimal generator
acts on the set of test functions (or functionals) $\phi:F\rightarrow\mathbb{R}$.
\black{The infinitesimal generator reads as}
\begin{equation}
G_{L}\left[\phi\right](n)=\lim_{\Delta\tau\downarrow0}\frac{\mathbb{E}_{n}\left[\phi(\hat{n}(\Delta\tau))\right]-\mathbb{\phi}(n)}{\Delta\tau}.\label{eq:Infinitetisimal_Generator}
\end{equation}
A key step to justify Eq. (\ref{eq:Path_Large_Deviation}) is
to prove it for an infinitesimal step. This amounts to computing
the probability that an increment $\left[\hat{n}(\Delta\tau)-n_{0}\right]/\Delta\tau$
is equal to a prescribed value denoted $\dot{n}$. If we can prove
that
\begin{equation}
\mathbb{P}_{n_{0}}\left[\frac{\hat{n}(\Delta\tau)-n_{0}}{\Delta\tau}=\dot{n}\right]\underset{L \to \infty}{\asymp}\exp\left(-\Delta\tau\left(\frac{L}{2\pi}\right)^{d}\sup_{\lambda}\left[\int\lambda(\bm{\xi}) \dot{n}(\bm{\xi}) \text{d}^d \xi-H[n_0,\lambda]\right]\right),\label{eq:Large_Deviation_Increment}
\end{equation}
then Eq. (\ref{eq:Path_Large_Deviation}) follows using the Markov property.
Here $\lambda$ appears as a variable conjugated to $\dot{n}$. As
explained in \cite[Sec. 7.1.2]{Bouchet2020}, if the limit
\begin{equation}
H\left[n,\lambda\right]=\lim_{L \to \infty}\left(\frac{2\pi}{L}\right)^{d} G_{L}\left[\mathrm{e}^{\left(\frac{L}{2\pi}\right)^{d}\int\text{d}^d \xi \,\lambda(\bm{\xi})\,\bullet}\right]\left[ n \right]\mbox{e}^{-\left(\frac{L}{2\pi}\right)^{d}\int\text{d}^d \xi \,\lambda(\bm{\xi})n(\bm{\xi})}\label{eq:H-Generator}
\end{equation}
exists, then Eq. (\ref{eq:Large_Deviation_Increment}) follows directly
from the definition of the infinitesimal generator, G\"artner-Ellis
theorem and simple computations. Then Eq. (\ref{eq:H-Generator}) justifies
the large deviation principle (\ref{eq:Path_Large_Deviation}). A
rigorous version of this simple explanation can be found in mathematical
textbooks \cite{feng2006large,FreidlinWentzell1998}, for some processes
with generic hypotheses.

For the weak turbulence problem of interest, we will proceed similarly.
However we have to adapt the reasoning in order to take into account
the  kinetic limit (\ref{eq:def_kinlim}).
To do so, let us consider the finite-time generator for the slow dynamics of
$n$, at a time step $\Delta\tau$:
\[
G_{L,\epsilon,\Delta\tau}\left[\phi\right]\left[n\right]=\frac{\mathbb{E}_{n}\left[\phi(\hat{n}\left(\cdot,\Delta\tau\right))\right]-\phi(n)}{\Delta\tau},
\]
where $\phi$ is a functional over the set of spectral density. \red{Contrary to the previous case, the limit $\Delta \tau \to 0$ cannot be taken before the kinetic limit (\ref{eq:def_kinlim}) since $\Delta \tau$ must satisfy (\ref{eq:kinetic_ineq}).} Consistently
with the discussion in section \ref{subsec:The-kinetic-limit}, the average
$\mathbb{E}_{n}$ is again a uniform probability distribution for
the phases (RP), conditioned on $\hat{n}(\cdot,0)=n$.
By analogy with (\ref{eq:H-Generator}), we define a large deviation
Hamiltonian $H$ by
\red{
\begin{equation}
  H[n,\lambda]=\lim_{\Delta \tau \to 0}\text{Kin}\lim \left(\frac{2\pi}{L}\right)^{d}G_{L,\epsilon,\Delta\tau}\left[\mathrm{e}^{\left(\frac{L}{2\pi}\right)^{d}\int\text{d}^d \xi \,\lambda(\bm{\xi})\,\bullet }\right]\left[n\right]\mathrm{e}^{-\left(\tfrac{L}{2\pi}\right)^{d}\int\text{d}^d \xi\,\lambda(\bm{\xi})n(\bm{\xi})},\label{eq:H-Kinetic-Definition}
\end{equation}
}
where we use the kinetic limit (\ref{eq:def_kinlim}) rather than
simply the $L\rightarrow\infty$ limit. \black{Then adapting the computations
in \cite[Sec. 7.1.2]{Bouchet2020}, using the G\"artner-Ellis
theorem, we conclude that the infinitesimal propagator for the empirical density is given by Eq. (\ref{eq:Large_Deviation_Increment}) where
the limit $L\rightarrow\infty$ is replaced by the kinetic limit. By iterating Eq. (\ref{eq:Large_Deviation_Increment}) (using the Markov property), we obtain (\ref{eq:Path_Large_Deviation}).}

Our goal is thus to compute (\ref{eq:H-Kinetic-Definition}). A simple
calculation shows that an equivalent formula for the large deviation
Hamiltonian is
\begin{equation}
H[n,\lambda] = \lim_{\Delta \tau \to 0}\frac{1}{\Delta\tau}  \text{Kin}\lim \left(\frac{2\pi}{L}\right)^{d} \log\mathbb{E}_{n}\left[\text{e}^{\left(\tfrac{L}{2\pi}\right)^{d}\int\text{d}^d \xi \,\lambda(\bm{\xi}) \left[ \hat{n}(\bm{\xi},\Delta \tau) - n(\bm{\xi})\right]}\right].\label{eq:def_Hamiltonian}
\end{equation}
\black{Therefore, it is helpful, in order to compute $H$, to define the moment generating function}
$Z_{L,\epsilon}$ of the empirical density increment $\hat{n}(\cdot,\Delta\tau)-n$. The latter is defined as
\begin{equation}
Z_{L,\epsilon}\left[n,\lambda,\Delta\tau\right]\equiv\mathbb{E}_{n}\left[{\rm e}^{\left(\frac{L}{2\pi}\right)^{d}\int\text{d}^d \xi \,\lambda(\bm{\xi})\left[\hat{n}(\bm{\xi},\Delta\tau)-n(\bm{\xi})\right]}\right].\label{eq:def_mgf}
\end{equation}
We then follow the classical approach to compute $Z_{L,\epsilon}$,
initiated by Peierls and followed by most of the classical literature
of weak turbulence, for instance \cite{Nazarenko2011}. Starting
from the evolution equations Eqs. (\ref{eq:bk_evol_eq}), one makes
a perturbative expansion at order two in $\epsilon$ for the
time evolution of $\left\{ b_{\bm{k}}\right\} $ up to a time $\Delta\tau\ll1$. One can then use this formula to perform explicitly the average $\mathbb{E}_{n}$ to compute (\ref{eq:def_mgf}). While the
technical aspects of these computations are very classical, and follow
the traditional approach, \black{our interpretation is slightly different since we condition on the value of $n$ and consider large deviations of the empirical spectrum. To our knowledge, such large deviation principle for the empirical spectrum in the kinetic limit cannot be found in the existing literature.} For the
sake of completeness, we perform explicitly the computation of $Z_{L,\epsilon}$
in Appendix \ref{sec:Appendix-B}. The result is
\begin{align}
& Z_{L,\epsilon}\left[n,\lambda,\Delta\tau\right]  \label{eq:perturbed_estimation_Z}\\
& \quad = 1 +\Delta\tau\left(\frac{L}{2\pi}\right)^{d}\left\{ 6\pi\sum_{\vec{\sigma}}\iiint{\rm d}^{d}\xi_{1}{\rm d}^{d}\xi_{2}{\rm d}^{d}\xi_{3}\left\lvert V_{\vec{\bm{\xi}}}^{\vec{\sigma}}\right\rvert^{2}\delta\left(\vec{\sigma}\cdot\vec{\bm{\xi}} \, \right)\delta\left(\vec{\sigma}\cdot\vec{\omega}\right)\right.\nonumber \\
 & \qquad\qquad\quad \times \left[\left(\vec{\sigma}\cdot\vec{\lambda}\right)\left(\sigma_{1}n(\bm{\xi}_{2})n(\bm{\xi}_{3})+\sigma_{2}n(\bm{\xi}_{1})n(\bm{\xi}_{3})+\sigma_{3}n(\bm{\xi}_{1})n(\bm{\xi}_{2})\right)\right. \nonumber \\
 & \qquad\qquad\qquad\left.\left.+\left(\vec{\sigma}\cdot\vec{\lambda}\right)^{2}n(\bm{\xi}_{1})n(\bm{\xi}_{2})n(\bm{\xi}_{3})\right]+\mathcal{R}_{1}(L,\epsilon)+\epsilon^{2}\mathcal{R}_{2}\left(L,\epsilon\right)\right\} \nonumber
\end{align}
with $\vec{\sigma}\cdot\vec{\lambda}=\sum_{i=1}^{3}\sigma_{i}\lambda(\bm{\xi}_{i})$.
The two terms $\mathcal{R}_{1}$ and $\mathcal{R}_{2}$ are two remainders
in the asymptotic expansion. This result can be compared to a slightly
different one in \cite[Eqs. (6.112-113), Chap. 6]{Nazarenko2011}.
Beyond the different interpretation, we also note that the prefactor
$\left(\frac{L}{2\pi}\right)^{d}$ does not appear in \cite[Eqs. (6.112-113), Chap. 6]{Nazarenko2011}.

The remainder $\mathcal{R}_{1}(L,\epsilon)$ corresponds to the approximation
of the discrete expression by continuous ones for all terms of order
$\epsilon^{2}$ in the expansion for the dynamics of $\left\{ b_{\bm{k}}\right\} $.
It is clear that $\text{Kin}\lim\mathcal{R}_{1}=0$. The second remainder
$\epsilon^{2}\mathcal{R}_{2}(L,\epsilon)$ is defined as the remainder term between the exact $\left\{ b_{\bm{k}}\right\}$ and its approximation up to order $2$ in $\epsilon$.
It is clear that $\epsilon^{2}\mathcal{R}_{2}(L,\epsilon)$ is of
order $\epsilon^{2}$ and that for fixed $L$, $\lim_{\epsilon\rightarrow0}\epsilon^{2}\mathcal{R}_{2}(L,\epsilon)=0$.
\black{However the actual dependence of $\epsilon^{2}\mathcal{R}_{2}(L,\epsilon)$
as $L$ increases is not controlled in 
our computation.
In the following we will simply assume
$\text{Kin}\lim\epsilon^{2}\mathcal{R}_{2}(L,\epsilon)=0$, and check that the result
is consistent with all the expected properties of the Hamiltonian}.

Assuming $\text{Kin}\lim\epsilon^{2}\mathcal{R}_{2}(L,\epsilon)=0$,
the connection between $Z_{L,\epsilon}$ \eqref{eq:perturbed_estimation_Z},
and the large deviation Hamiltonian $H$ \eqref{eq:def_Hamiltonian}
is easily understood by expanding the logarithm. \red{After a straightforward reorganisation of the terms in \eqref{eq:perturbed_estimation_Z}, one finally obtains the large
deviation Hamiltonian
\begin{align}
H[n,\lambda] & =6\pi\sum_{\vec{\sigma}}\iiint{\rm d}^{d}\xi_{1}{\rm d}^{d}\xi_{2}{\rm d}^{d}\xi_{3}\left\lvert V^{\vec{\sigma}}_{\vec{\bm{\xi}}}\right\rvert^{2}\delta\left(\vec{\sigma}\cdot\vec{\bm{\xi}}\,\right)\delta\left(\vec{\sigma}\cdot\vec{\omega}\right) \label{eq:H_final_result} \\
             & \qquad\qquad \times \left(\vec{\sigma}\cdot\vec{\lambda}\right)n(\bm{\xi}_{1})n(\bm{\xi}_{2})n(\bm{\xi}_{3}) \left[ \vec{\sigma}\cdot\left(\frac{1}{\vec{n}}+\vec{\lambda} \right)\right]
               \nonumber
\end{align}
with the notation $1/\vec{n}=(n^{-1}(\bm{\xi}_1),n^{-1}(\bm{\xi}_2),n^{-1}(\bm{\xi}_3))$.}

\black{This derivation casts the large-deviations theory for wave kinetics in the same mathematical
framework as that for the Boltzmann equation of low-density gases \cite{Bouchet2020},
for the Landau equation of weakly-coupled plasmas \cite{feliachi2021dynamical}, or for the Lenard-Balescu equation of particle systems with long range interactions \cite{feliachi2022dynamical}.}

\black{It is worth mentioning, however, that there is another derivation of this same result that is even
closer to the derivation of the traditional Peierls equation. This alternative derivation proceeds by defining an
unconditional generating functional
\begin{equation}
Z_{L,\epsilon}[\lambda,\tau]=\mathbb{E}\left[\exp\left( \left(\frac{L}{2\pi}\right)^d \int \lambda(\bm{\xi}) \hat{n}(\bm{\xi},\tau) \right)\right]
= \mathbb{E}\left[\exp\left( \sum_{\bm{k}} \lambda(\bm{k}) \vert b_{\bm{k}}(\tau)\vert^2 \right)\right]
\end{equation}
which is the same quantity that appears in the Peierls equation, except for the normalization
of the expression in the exponent. A main assumption of this approach is that a {\it free-energy functional}
exists which is defined by the kinetic limit:
\begin{equation}
F[\lambda,\tau]=\text{Kin}\lim \left(\frac{2\pi}{L}\right)^d \ln Z_{L,\epsilon}[\lambda,\tau] \label{eq:Fdef}
\end{equation}
which implies that a large-deviations property holds for the empirical spectral density
\eqref{eq:def_empirical_spectrum} at each instant of macroscopic time $\tau,$ with a rate function which
is given by the Legendre transform
\begin{equation}
I[n,\tau]=\sup_\lambda\left\{ \int \lambda(\bm{\xi})n(\bm{\xi})\mathrm{d}^{d}\xi -F[\lambda,\tau] \right\}.
\end{equation}
There is no need to condition upon
deterministic initial data $n_0(\bm{k})$ in this approach, but instead the limit \eqref{eq:Fdef} is
assumed to exist also at time $t=0$ so that a large-deviations property holds initially for the
empirical spectral density. Then, defining the time derivative, with $F_{L,\epsilon}[\lambda,\tau]:=
\left(\frac{2\pi}{L}\right)^d \ln Z_{L,\epsilon}[\lambda,\tau],$ by
\red{
\begin{align*}
\frac{\partial F[\lambda]}{\partial\tau}
 & = \lim_{\Delta \tau \to 0} \frac{1}{\Delta\tau} \text{Kin}\lim \left(F_{L,\epsilon}[\lambda,\Delta\tau]-F_{L,\epsilon}[\lambda,0]\right)  \\
 & =\lim_{\Delta \tau \to 0} \frac{1}{\Delta\tau} \text{Kin}\lim\left(\frac{2\pi}{L}\right)^{d}
\frac{Z_{L,\epsilon}[\lambda,\Delta\tau]-Z_{L,\epsilon}[\lambda,0]}{Z_{L,\epsilon}[\lambda,0]}
\end{align*}
}
a calculation similar to the preceding one yields the equation
\begin{equation}
\frac{\partial F[\lambda]}{\partial\tau} = H\left[\frac{\delta F}{\delta\lambda},\lambda\right]
\label{Feq}
\end{equation}
where $H[n,\lambda]$ is the large deviation Hamiltonian \eqref{eq:H_final_result}. Since
$\partial I/\partial\tau=-\partial F/\partial\tau,$ an equivalent {\it Hamilton-Jacobi equation}
\begin{equation}
\frac{\partial I[n]}{\partial\tau} + H\left[n,\frac{\delta I}{\delta n}\right] =0
\end{equation}
holds for the single-time rate function $I[n,\tau].$ As familiar from classical mechanics, this equation
may be solved by the method of characteristics, yielding the least-action formula
\begin{equation}
I[n,\tau]=\inf_{\{\bar{n}:\, \bar{n}(\tau)=n,\, \bar{n}(0)=n_0\}} \{I[n_0,0]+\mathcal{A}[\bar{n}]\}
\end{equation}
with the action $\mathcal{A}[n]$ defined in Eq.\eqref{eq:stochastic_action}. This is exactly the relation
required by the Contraction Principle of large deviations theory.}

The kinetic equation naturally appears as the equation generating
the most probable path for the empirical spectral density $\hat{n}$.
The latter is obtained by minimizing the stochastic action \eqref{eq:stochastic_action},
which yields $\dot{n}\left(\bm{\xi},\tau\right)=\frac{\delta H}{\delta\lambda(\bm{\xi})}\left[n,0\right]$.
Computing this expression, we indeed obtain the expected kinetic equation
\begin{equation}
  \label{eq:kinetic_eq}
  \dot{n}\left(\bm{\xi},\tau\right) = \mathcal{C}[n](\bm{\xi})
\end{equation}
\black{with $\mathcal{C}[n]$ the collision integral that can be split into self-consistent forcing
and damping rate terms
\begin{equation}
\mathcal{C}[n](\bm{\xi}) = \eta(\bm{\xi},\tau)-\gamma(\bm{\xi},\tau)n(\bm{\xi},\tau)
\end{equation}
with
\begin{align}
   \label{selfcons}
\eta(\bm{\xi},\tau) &= 36\pi\sum_{\substack{\vec{\sigma}\\\sigma_{1}=-1}
}\iint{\rm d}^{d}\xi_{2}{\rm d}^{d}\xi_{3}\left\lvert V_{\vec{\bm{\xi}}}^{\vec{\sigma}}\right\rvert^{2}\delta\left(\vec{\sigma}\cdot\vec{\bm{\xi}} \, \right)\delta\left(\vec{\sigma}\cdot\vec{\omega}\right) n(\bm{\xi}_{2})n(\bm{\xi}_{3})  \\
\gamma(\bm{\xi},\tau) &= 36\pi\sum_{\substack{\vec{\sigma}\\\sigma_{1}=-1}
}\iint{\rm d}^{d}\xi_{2}{\rm d}^{d}\xi_{3}\left\lvert V_{\vec{\bm{\xi}}}^{\vec{\sigma}} \right\rvert^{2}\delta\left(\vec{\sigma}\cdot\vec{\bm{\xi}} \, \right)\delta\left(\vec{\sigma}\cdot\vec{\omega}\right)
  \left[\sigma_{2}n(\bm{\xi}_{3})+\sigma_{3}n(\bm{\xi}_{2})\right] \nonumber
\end{align}
}

The large deviation Hamiltonian $H$ in \eqref{eq:H_final_result} is quadratic in the response
field $\lambda$ which means that the statistics of the local time increments
$\dot{n}\mathrm{d}\tau$ is Gaussian. \black{Formally,
\eqref{eq:H_final_result} has the form of a Freidlin-Wentzell large-deviations Hamiltonian
for a weak-noise diffusion process
\begin{equation}
 H[n,\lambda]= \langle \mathcal{C}\left[ n \right],\lambda\rangle+\langle\lambda,\mathcal{Q}\left[ n \right]\lambda\rangle
 \label{FWHam}
 \end{equation}
where $\mathcal{C}\left[ n \right]$ is the collision integral defined in \eqref{eq:kinetic_eq} and $\mathcal{Q}\left[ n \right]$
is a nonnegative-definite, self-adjoint operator  which may  be interpreted as a ``noise covariance'',
associated to the quadratic form
\begin{align}
\langle\lambda,\mathcal{Q}\left[ n \right]\lambda\rangle& =
6\pi\sum_{\vec{\sigma}}\iiint{\rm d}^{d}\xi_{1}{\rm d}^{d}\xi_{2}{\rm d}^{d}\xi_{3}\left\lvert V_{\vec{\bm{\xi}}}^{\vec{\sigma}}\right\rvert^{2}\delta\left(\vec{\sigma}\cdot\vec{\bm{\xi}}\right)\delta\left(\vec{\sigma}\cdot\vec{\omega}\right) \label{quad} \\
   & \qquad \qquad \times \left(\vec{\sigma}\cdot\vec{\lambda}\right)^{2} n(\bm{\xi}_{1})n(\bm{\xi}_{2})n(\bm{\xi}_{3}) \nonumber
\end{align}
A straightforward calculation shows that the kernel of this operator is
\begin{align}
 \mathcal{Q}\left[ n \right](\bm{\xi},\bm{\xi}') & = \eta(\bm{\xi})n(\bm{\xi})\delta(\bm{\xi}-\bm{\xi}') \label{Qdef}  \\
& \quad + 36\pi\sum_{\vec{\sigma}}\int{\rm d}^{d}\xi_{3}\left\lvert V_{\vec{\bm{\xi}}}^{\vec{\sigma}}\right\rvert^{2}
\delta\left(\vec{\sigma}\cdot\vec{\bm{\xi}}\right)\delta\left(\vec{\sigma}\cdot\vec{\omega}\right) n(\bm{\xi}_{3}) \cdot n(\bm{\xi})n(\bm{\xi}') \nonumber
\end{align}
where the part delta-correlated in wavenumber is proportional to the  self-consistent forcing. Because of this structure,
the Legendre transform of $H[n,\lambda]$ has the form of an Onsager-Machlup Lagrangian
\begin{equation}
L[n,\dot{n}]=\frac{1}{4} \iint {\rm d}^{d}\xi \, {\rm d}^{d}\xi'\
(\dot{n}(\bm{\xi})-\mathcal{C}[n](\bm{\xi})) \cdot  \mathcal{Q}\left[ n \right]^{-1}(\bm{\xi},\bm{\xi}') \cdot (\dot{n}(\bm{\xi}')-\mathcal{C}[n](\bm{\xi}'))
\label{OM-lag} \end{equation}
Note that here $\mathcal{Q}\left[ n \right]^{-1}(\bm{\xi},\bm{\xi}')$ is the kernel of the operator pseudo-inverse, since the
quadratic form \eqref{quad} is degenerate, vanishing whenever $\lambda(\bm{\xi})=\bm{\mu}\cdot\bm{\xi}+ \beta\cdot \omega(\bm{\bm{\xi}})$
for any constants $\bm{\mu},$ $\beta.$ This is related to symmetry properties of the large-deviations Hamiltonian,
discussed in the following section.}

\red{
\subsection{Beyond $3$-waves interactions: higher-order nonlinearities}\label{higher}

For the sake of simplicity, the microscopic Hamiltonian (\ref{eq:H_dynamics}) that we have considered in this paper involves 3-wave interactions only. Nonetheless, the large deviation Hamiltonian (\ref{eq:H_final_result}) can also be obtained for nonlinear wave systems with higher-order nonlinearities with the very same hypothesis (provided that the nonlinear term of the equation of motion (\ref{eq:ak_evol_eq}) scales as $\epsilon L^{-d/2}$).
\red{An important case is the one with 4-wave interactions, discussed in \cite[section 2.1.5]{Zakharov1992},  \cite[section 6.9]{Nazarenko2011}, or,
most similar to our present treatment \cite{chibbaro2018, chibbaro2017wave,Shi2016}. This case appears for instance in the nonlinear Schr\"{o}dinger dynamics \cite{Dyachenko1992}, deep-water gravity waves \cite{zakharov1966energy} or vibrations of elastic plates (F\"{o}ppl-von K\'{a}rm\'{a}n equation) \cite{landau1986theory}.}
Following \cite{chibbaro2018}, a generic microscopic Hamiltonian for 4-wave interactions reads as
\begin{equation}
  \mathcal{H} = \sum_{\bm{k}} \omega_{\bm{k}} {\left\lvert A_{\bm{k}} \right\rvert}^{2} + \sum_{\vec{\sigma}} \sum_{\vec{\bm{k}}} W_{\vec{\bm{k}}}^{\vec{\sigma}} A_{\bm{k}_1}^{\sigma_1} A_{\bm{k}_2}^{\sigma_2} A_{\bm{k}_3}^{\sigma_3} A_{\bm{k}_4}^{\sigma_4}
  \delta_{\vec{\sigma}\cdot\vec{\bm{k}},0} \, ,
  \label{eq:4-waves-Hamiltonian}
\end{equation}
with $\vec{\sigma}=(\sigma_1,\sigma_2,\sigma_3,\sigma_4)$ ($\sigma_i=\pm 1$) and $\vec{\bm{k}}=(\bm{k}_1, \bm{k}_2, \bm{k}_3, \bm{k}_4)$. We assume that the interaction kernel $W$ satisfies the conditions (\ref{eq:prop_V_hamiltonian}).

Following the same steps detailed in Appendix \ref{sec:Appendix-B}, we anticipate that the corresponding large deviation Hamiltonian reads as
\begin{align}
  & H[n,\lambda] = 24 \pi \sum_{\vec{\sigma}} \int \mathrm{d}^{d}\xi_1 \mathrm{d}^{d}\xi_2 \mathrm{d}^{d}\xi_3 \mathrm{d}^{d}\xi_4 \,  { \left\lvert W_{\vec{\bm{\xi}}}^{\vec{\sigma}} \right\rvert }^{2}\delta(\vec{\sigma}\cdot\vec{\bm{\xi}}) \delta(\vec{\sigma}\cdot\vec{\omega}) \label{eq:LD_hamiltonian_4waves_interaction} \\
  & \qquad \qquad  \times \left( \vec{\lambda}\cdot\vec{\sigma} \right) n(\bm{\xi}_1)n(\bm{\xi}_2)n(\bm{\xi}_3)n(\bm{\xi}_4) \left[ \vec{\sigma}\cdot\left(\frac{1}{\vec{n}} + \vec{\lambda}\right) \right] \; . \nonumber
\end{align}
Note that the frequency renormalization often considered in 4-wave interacting systems \cite{chibbaro2018, chibbaro2017wave,Nazarenko2011} has been neglected here since the latter appears at first order in $\epsilon$ in the kinetic limit.

The large deviation Hamiltonian (\ref{eq:LD_hamiltonian_4waves_interaction}) has the very same structure as the $3$-wave one (\ref{eq:H_final_result}). Therefore, all the generic properties of the large deviation dynamics for the $3$-wave interaction system (see next section \ref{sec:hamiltonian_properties}) are recovered for higher-order nonlinearities. The 4-wave system has just an extra conserved quantity that is the total wave action (in addition to the energy and the momentum).
}

\section{Properties of the large deviation Hamiltonian and equilibrium quasipotential}\label{sec:hamiltonian_properties}

We now investigate the symmetry properties of the large deviation
Hamiltonian $H$ \eqref{eq:H_final_result} associated to conservation
laws. In the specific case of wave dynamics truncated at a finite
number of modes, we also check that the Hamiltonian $H$ is compatible
with the quasipotential for the microcanonical measure, and has a
symmetry associated to time-reversal symmetry.

\subsection{Conservation laws (for $3$-wave interactions)}\label{subsec:Conservation_properties}

As explained in \cite[Sec. 7.2.6]{Bouchet2020}, each conservation
law of the equations is associated to a symmetry of the Hamiltonian
$H$. If $C[n]$ is a conserved quantity for any evolution of the
empirical spectrum $n$, then
\begin{equation}
H\left[n,\lambda+\alpha\frac{\delta C}{\delta n}\right]=H\left[n,\lambda\right]\label{eq:conservation_law_large_dev}
\end{equation}
for every conjugated field $\lambda$ and $\alpha \in \mathbb{R}$.

\subsubsection{Energy conservation}

We first consider energy conservation. The Hamiltonian dynamics
(\ref{eq:ak_evol_eq}-\ref{eq:bk_evol_eq}) conserves the total microscopic energy $\mathcal{H}=\mathcal{H}_{2}+\mathcal{H}_{3}$.
The second order term can be expressed exactly as a function of the
empirical density, namely $\mathcal{H}_{2}=E[n]$ with
\[
E[n]=\int{\rm d}^{d}\xi\:\omega_{\bm{\xi}}n(\bm{\xi}).
\]
The energies $\mathcal{H}_{2}$ and $\mathcal{H}_{3}$ are not independently
conserved by the Hamiltonian dynamics, only $\mathcal{H}$ is. However,
our derivation assumes that the quadratic term $\mathcal{H}_{2}$
dominates the cubic term: $\mathcal{H}_{3}/\mathcal{H}_{2} = O(\epsilon)$ by assumption. This hypothesis prevents exchanges of energy between
$\mathcal{H}_{2}$ and $\mathcal{H}_{3}$ which are of order larger
than $\epsilon$. As a consequence we can deduce that
\[
\mathrm{Kin\, lim}\frac{E\left[\hat{n}(\Delta \tau)\right]-E\left[\hat{n}(0)\right]}{\Delta \tau} =0.
\]
In terms of probability, one gets:
\[
\lim_{\Delta \tau \to 0}\text{Kin}\lim\mathbb{P}_{n_{0}}\left(\frac{E\left[\hat{n}\left(\Delta\tau\right)\right]-E\left[\hat{n}(0)\right]}{\Delta\tau}=\dot{E}\right)=\delta\left(\dot{E}\right).
\]
We thus conclude that $E\left[n\right]$ must be conserved
by the weak-noise (kinetic limit) dynamics.

\paragraph{}

We can check this property directly. Writing, $\tfrac{\delta E}{\delta n(\bm{\xi})}=\omega_{\bm{\xi}}$,
it is straightforward to verify the symmetry $H\left[n,\lambda+\alpha\frac{\delta E}{\delta n}\right]=H\left[n,\lambda\right]$,
from (\ref{eq:H_final_result}). This is a simple consequence of the
triad energy constrain $\vec{\sigma}\cdot\vec{\omega}=0$ in the integral.

\subsubsection{Momentum conservation}

The momentum
$$
\bm{K}=\sum_{\bm{k}\in\mathbb{Z}_{L}^{d}} \bm{k} \left\lvert a_{\bm{k}}\right\rvert^{2} = \int \! \bm{\xi} \, \hat{n}(\bm{\xi}) {\rm d}^{d}\xi
$$
is exactly conserved by the microscopic dynamics \eqref{eq:bk_evol_eq}.
Hence it must be also conserved at the level of the large deviations
of the empirical spectrum. The symmetry $H\left[n,\lambda+\bm{\alpha} \cdot \frac{\delta \bm{K}}{\delta n}\right]=H\left[n,\lambda\right]$
is easily verified because of the presence of the constrain $\vec{\sigma}\cdot\vec{\bm{k}}=0$
in the Hamiltonian $H$ (\ref{eq:H_final_result}).

\paragraph{}

\black{
  To conclude, we would like to emphasize that we have only considered generic conservation laws for $3$-wave interacting systems here. We notice that other conservation laws may be present. For instance, the Kadomtsev-Petviashvili equation conserves as well a third quantity called the zonostrophy \cite[Sec. 8.1.2.1]{Nazarenko2011}.}
\black{Such ``emergent'' conservation laws will not generally be respected by the large deviations.}

\subsection{Quasipotential and detailed balance for the microcanonical measure}

The stochastic action $\mathcal{A}[n]$ in \eqref{eq:stochastic_action}
quantifies the probability to observe trajectories of the \black{weak-noise} stochastic
process $\hat{n}(\cdot,\tau)$.

In the kinetic limit, the stochastic process $\hat{n}(\cdot,\tau)$
becomes deterministic and follows the relaxation dynamics as described
by the kinetic equation \eqref{eq:kinetic_eq}. In equilibrium, for
a given total energy $E$ and momentum $\bm{K}$ (that are conserved,
as seen in Sec. \ref{subsec:Conservation_properties}) , one expects
the spectral density $n(\cdot,\tau)$ to relax toward a unique stationary
solution (\emph{i.e.} attractor) $n^{\ast}$ of the kinetic equation
\eqref{eq:kinetic_eq}. The latter is known to be the Rayleigh-Jeans
spectrum in equilibrium \cite{Nazarenko2011} and reads as
\begin{equation}
n^{\ast}(\bm{\xi})=\left[\beta\omega_{\bm{\xi}}+\bm{\mu}\cdot\bm{\xi}\right]^{-1}\label{eq:RJ_spectrum}
\end{equation}
where $\beta$ and $\bm{\mu}$ are Lagrange parameters that are fixed
by the constrains $E=\int{\rm d}^{d}\xi\ \omega_{\bm{\xi}}n^{\ast}(\bm{\xi})$
and $\bm{K}=\int{\rm d}^{d}\xi\ \bm{\xi}\, n^{\ast}(\bm{\xi})$\footnote{Note that the Rayleigh-Jeans spectrum as defined in (\ref{eq:RJ_spectrum})
for the full Fourier space $\mathbb{R}^{d}$ yields ultraviolet
divergence \cite[Chap. 9]{Nazarenko2011}. In order to make sense
of it, one has to restrict the allowed Fourier space by introducing a UV
cut-off $k_{{\rm max}}$. This will be briefly discussed in the following
subsection and Appendix \ref{sec:Appendix-Quasipotential_and_Entropy}.}.

For finite $L$, fluctuations are present and one observes noise and
rare excursions around the deterministic trajectory. In the stationary
regime, fluctuations are quantified by the stationary probability
density $P_{L,\epsilon}[n]$, whose maximum is reached at $n=n^{\ast}$.

In the kinetic limit, the distribution $P_{L,\epsilon}[n]$ is characterised
by a large deviation rate function
\begin{equation}
Q[n]=-\text{Kin lim}\left(\frac{2\pi}{L}\right)^{d} \ln P_{L,\epsilon}[n]\,,\label{eq:def_quasipotential-1}
\end{equation}
called the quasipotential. The latter quantifies the (rare) fluctuations
in the kinetic limit ($L\to\infty$, $\epsilon\to 0$) around the stationary
solution(s) $n^{\ast}$ of the Kinetic equation (\ref{eq:kinetic_eq}).
The quasipotential $Q[n]$ \black{is the special case of the instantaneous rate-function $I[n,\tau]$
in the limit $\tau\to\infty$ and can be computed from the solutions
of the stationary Hamilton-Jacobi equation $H[n,\frac{\delta Q}{\delta n}]=0$.}

However, since the microscopic dynamics considered here is Hamiltonian
(no driving force, no dissipation), one can also rely on the principle
of equilibrium statistical mechanics to compute the quasipotential
$Q$ at equilibrium.

\subsubsection{Quasipotential at equilibrium from the microcanonical distribution}

The microscopic dynamics (\ref{eq:ak_evol_eq}) (finite $L$, finite
$\epsilon$) is a Hamiltonian dynamics that derives from the Hamiltonian
(\ref{eq:H_dynamics}). The microscopic energy $E=\mathcal{\tilde{H}}\equiv\epsilon^{-2}\mathcal{H}$
(which has been rescaled to be expressed in terms of the modes $a_{\bm{k}}$)
is thus conserved by definition of the dynamics. Furthermore, we have
seen previously that the momentum $\bm{K}=\sum_{\bm{k}}\bm{k}\vert a_{\bm{k}}\vert^{2}$
is also conserved by the microscopic dynamics.

Until now, we have considered the full Fourier space $\mathbb{Z}_L^{d}$.
Our expressions are thus valid as long as the modes
$\{a_{\bm{k}}\}$ decay sufficiently fast with $\bm{k}$ such that
the sums are convergent. However, when considering the microcanonical
distribution over the modes $\left\{ a_{\bm{k}}\right\} $ or the
associated equilibrium (Rayleigh-Jeans) spectrum $n^{\ast}(\bm{k})$
(\ref{eq:RJ_spectrum}), one will have to restrict the $\bm{k}$-space
by introducing the bounded set $\mathbb{K}_{L}^{d}=\left\{ \bm{k}=\mathbb{Z}_L^{d}\left\lvert \left\vert  \bm{k}\right\vert  \leqslant k_{{\rm max}}\right.\right\} $
in order to avoid divergences. The dynamics of the modes can be restricted
to the space $\left\vert  \bm{k}\right\vert  \leqslant k_{{\rm max}}$
by setting $V_{\bm{k}_{1},\bm{k}_{2},\bm{k}_{3}}^{\sigma_{1},\sigma_{2},\sigma_{3}}=0$
for any $\left\vert  \bm{k}_{i}\right\vert  >k_{{\rm max}}$, $i=1,2,3$.
One thus avoids transport of energy beyond a certain threshold $k_{{\rm max}}$.
In this section, all the sums and the product over the wavenumbers
$\bm{k}$ are implicitly restricted to $\mathbb{K}_{L}^{d}$.

Let us now introduce the microcanonical distribution. One checks that
the flat measure $\prod_{\bm{k}}{\rm d}a_{\bm{k}}{\rm d}a_{\bm{k}}^{\ast}=\prod_{\bm{k}}{\rm d}\vert a_{\bm{k}}\vert^{2}{\rm d}\varphi_{\bm{k}}$
is stationary along the trajectories of the microscopic dynamics.
Hence, according to the microcanonical principle of equilibrium statistical
mechanics, the microcanonical measure associated with the macrostate
of energy $E$ and momentum $\bm{K}$ reads as
\begin{equation}
\text{d}\mu_{E,\bm{K},L,\epsilon}=\frac{1}{\Gamma_{E,\bm{K},L,\epsilon}}\delta\left(E-\mathcal{\tilde{H}}\right)\delta\left(\bm{K}-\left(\tfrac{2\pi}{L}\right)^{d}\sum_{\bm{k}}\bm{k}\left\lvert a_{\bm{k}}\right\rvert^{2}\right)\prod_{\bm{k}}{\rm d}a_{\bm{k}}{\rm d}a_{\bm{k}}^{\ast},\label{eq:microcanonical_distribution-2}
\end{equation}
with $\Gamma_{E,\bm{K},L,\epsilon}(E,\bm{K})=\prod_{\bm{k}}\int{\rm d}a_{\bm{k}}{\rm d}a_{\bm{k}}^{\ast}\delta\left(E-\tilde{\mathcal{H}}\right)\delta\left(\bm{K}-\left(\tfrac{2\pi}{L}\right)^{d}\sum_{\bm{k}}\bm{k}\left\lvert a_{\bm{k}}\right\rvert^{2}\right)$
the volume of the phase space associated with the macrostate $(E,\bm{K})$.

Defining the microcanonical probability distribution $P_{E,\bm{K},L,\epsilon}[n]=\mathbb{E}_{E,\bm{K},L,\epsilon}\left[\delta\left(\hat{n}-n\right)\right]$,
with $\mathbb{E}_{E,\bm{K},L,\epsilon}$ denoting the expectation with
respect to the microcanonical measure (\ref{eq:microcanonical_distribution-2}),
we show in Appendix \ref{sec:Appendix-Quasipotential_and_Entropy}
that the quasipotential $Q_{E,\bm{K}}[n]=-\text{Kin Lim}\left(\frac{2\pi}{L}\right)^{d}\log P_{E,\bm{K},L,\epsilon}[n]$,
associated with the microcanonical distribution reads \newpage
\begin{align}
& Q_{E,\bm{K}}[n] = \label{eq:microcanonical_quasipotential} \\
& \: \begin{cases}
\int{\rm d}^{d}\xi\,\left[\frac{n(\bm{\xi})}{n^{\ast}(\bm{\xi})}-
\log\left(\frac{n(\bm{\xi})}{n^{\ast}(\bm{\xi})}\right)-1\right] & \text{if }E=\int{\rm d}^{d}\xi\,\omega_{\bm{\xi}}n(\bm{\xi})\,,\;\bm{K}=\int{\rm d}^{d}\xi\,\bm{\xi}n(\bm{\xi})\\
+\infty & \text{otherwise}
\end{cases}. \nonumber
\end{align}
with $n^{\ast}(\bm{\xi})$ the equilibrium Rayleigh-Jeans spectrum (\ref{eq:RJ_spectrum}).

We check that this large deviations function $Q_{E,\bm{K}}$ indeed
satisfies the stationary Hamilton\textendash Jacobi equation $H\left[n,\delta Q/\delta n\right]=0$
and is thus admissible to be the quasipotential of the large deviation
dynamics of $\hat{n}$.

\subsubsection{Detailed balance property at equilibrium}

Since the microscopic dynamics \eqref{eq:ak_evol_eq} is Hamiltonian,
it is symmetric with respect to time inversion. The equations of motion
\eqref{eq:ak_evol_eq} are precisely symmetric with respect to the transformation
$a_{\bm{k}}^{(R)}(t)=a_{\bm{k}}^{\ast}(t_{{\rm fin}}-t)$ for $\left\{ a_{\bm{k}}(t)\right\} _{0\leqslant t\leqslant t_{{\rm fin}}}$a
trajectory of length $t_{{\rm fin}}$. In terms of the spectral density
$n$, the time-reversal symmetry is simply $n^{(R)}(t)=n(t_{{\rm fin}}-t)$.
Although time-reversal symmetry is generally lost when looking only
at the relaxation dynamics (\emph{e.g. }the dynamics as described
by the Kinetic equation), we show that the latter is restored when
fluctuations (large deviations) are present.

If $Q$ refers to the quasipotential of the dynamics, the time-reversal
symmetry of the stochastic action $\mathcal{A}$ is equivalent to the property:
\begin{equation}
H\left[n,\lambda+\frac{\delta Q}{\delta n}\right]=H\left[n,-\lambda\right]\label{eq:LD_detailed_balance}
\end{equation}
for every $\lambda$ and $n$. \red{A proof of this property is provided in \cite[Sec. 7.3.1]{Bouchet2020}.}
The symmetry relation \eqref{eq:LD_detailed_balance}
is referred to as the large deviation detailed balance relation.

We easily check that the large deviation detailed balance is verified
at equilibrium with respect to the quasipotential $Q=Q_{E,\bm{K}}$ (\ref{eq:microcanonical_quasipotential})
derived from the microcanonical distribution.

\section{Comparison with independent modes interacting in a mean-field way}

\subsection{Definition of the stochastic mean-field dynamics and associated Hamiltonian}

The Hamiltonian \eqref{eq:H_final_result} looks very much like the
Hamiltonian of a mean-field system made of modes that would only interact
through the global empirical spectral density $\hat{n}$.

Indeed, let us consider $\mathcal{N}_{L}$ ($\propto\left(\frac{L}{2\pi}\right)^{d})$
variables $J_{\bm{k}} > 0$ ($\bm{k}\in\mathbb{Z}_L^{d}$)
that interact through a mean-field coupling according to the following
Langevin dynamics (It\={o}):
\begin{equation}
\mathrm{d}J_{\bm{k}}=\left(\eta[\hat{n}](\bm{k})-J_{\bm{k}}\gamma[\hat{n}](\bm{k})\right)\mathrm{d}t+\sqrt{2J_{\bm{k}}\eta[\hat{n}](\bm{k})}\mathrm{d}W_{\bm{k}}\label{eq:mean-field_dynamics}
\end{equation}
with $\hat{n}:\bm{\xi}\mapsto\frac{1}{\mathcal{N}_{L}}\sum_{\bm{k}}J_{\bm{k}}\delta(\bm{\xi}-\bm{k})$
the empirical spectral density and $\mathrm{d}W_{\bm{k}}$ a Gaussian
white noise with variance $\mathbb{E}\left[\mathrm{d}W_{\bm{k}_{1}}(t)\mathrm{d}W_{\bm{k}_{2}}(t')\right]=\delta_{\bm{k}_{1},\bm{k}_{2}}\delta(t-t')
\mathrm{d}t.$
\black{The functionals $\gamma(\bm{k})$ and $\eta(\bm{k})$ appearing in \eqref{eq:mean-field_dynamics}
were defined in \eqref{selfcons}.}
This equation corresponds to the dynamics
of the square amplitudes of mode $\bm{k}$, namely $J_{\bm{k}}=\left\lvert a_{\bm{k}}\right\rvert^{2}$,
coupled through the spectrum $\hat{n}$ (that becomes non-fluctuating
in the kinetic limit). Its deterministic evolution
is precisely the one prescribed by the kinetic equation \eqref{eq:kinetic_eq}.
It is directly inspired by the study of the one-mode statistics within
the \emph{Random Phase }and \emph{Amplitude }(RPA) approximation obtained
in the kinetic limit \cite[Chap. 6]{Nazarenko2011} \black{
and it was previously shown in \cite{EyinkShi2012} to reproduce the exact
evolution equation for such one-mode statistics.}
Here, we go one step backward and define a mean-field system for finite
$L$, for which the RPA approximation is broken.

One can proceed in a similar way as for the original dynamics of the
modes done previously [we refer to \cite{feliachi2021dynamical}
where the detailed calculation is expounded]. The large deviations
Hamiltonian (\ref{eq:def_Hamiltonian}) associated
with the fluctuations of the empirical spectrum $\hat{n}$ in this mean-field system reads as
\begin{align}
H_{{\rm MF}}[n,\lambda] & =\int{\rm d}^{d}\bm{\xi}\:\left\{ \lambda(\bm{\xi})\left(\eta[n](\bm{\xi})-n(\bm{\xi})\gamma[n](\bm{\xi})\right)+\lambda(\bm{\xi})^{2}n(\bm{\xi})\eta[n](\bm{\xi})\right\} \,.\label{eq:H_mean-field}
\end{align}

\subsection{Comparison of both Hamiltonians (\ref{eq:H_mean-field}) and (\ref{eq:H_final_result})}

Both Hamiltonians $H$ (\ref{eq:H_final_result}) and $H_{{\rm MF}}$
(\ref{eq:H_mean-field}) describe fluctuations of \black{weak-noise diffusive systems}. The linear terms in $\lambda$ are the same.
Hence, the two Hamiltonians yield both the same Kinetic equation (\ref{eq:kinetic_eq}) and thus describe the very same relaxation dynamics of the spectral density $n$. Their difference lies in the quadratic
term in $\lambda$ that represents the correlations of the Gaussian
current $\dot{n}{\rm d}\tau$. \black{Although $H_{{\rm MF}}$ has the same Freidlin-Wentzell form \eqref{FWHam} as the true large-deviations Hamiltonian
of wave-kinetics $H$ in \eqref{eq:H_final_result}, it retains only the delta-correlated part of the noise covariance
$\mathcal{Q}[n]$ in \eqref{Qdef} and it is missing the part off-diagonal in wavenumber.}

\red{To emphasize the difference between the two theories, we  can formulate our new
Hamiltonian (\ref{eq:H_final_result}) as a formally equivalent nonlinear Langevin model
for the empirical spectrum, with non-local, multiplicative noise
\begin{equation} 
\mathrm{d} \hat{n}(\bm{\xi})={\mathcal C}[\hat{n}](\bm{\xi}) \mathrm{d}\tau + \left(\frac{2\pi}{L}\right)^{d/2}
    \int \mathrm{d}^d\bm{\xi}' \,\, (2Q)^{1/2}[\hat{n}](\bm{\xi},\bm{\xi}') \mathrm{d}W(\bm{\xi}',\tau),  \label{LD-Langevin}
\end{equation}
where $\mathrm{d}W(\bm{\xi},\tau)$ is a Gaussian white-noise field of mean zero and covariance
$\mathbb{E}\left[\mathrm{d}W(\bm{\xi},\tau) \mathrm{d}W(\bm{\xi}',\tau')\right] =\delta^d(\bm{\xi}-\bm{\xi}')\delta(\tau-\tau')
\mathrm{d}\tau$ and where $Q^{1/2}$ is any square root of the operator $Q$ with kernel
\eqref{Qdef}.}
For the mean-field system, the modes become virtually independent in the kinetic limit. This asymptotic
independence of two distinct modes translates into the $\delta$-correlation (in terms of the wavenumbers
$\bm{k}$) of the variance of the noise. On the other hand, large deviations of the spectral density within
the RP assumption yields \red{the stochastic model \eqref{LD-Langevin} driven by} a Gaussian noise
with non-trivial covariance $\propto Q$ that couples together modes with distinct wavenumbers $\bm{\xi}.$

The latter coupling does not affect the mean relaxation dynamics described by the Kinetic equation.
\red{Indeed, the prior predictions \cite{lvov2004noisy} for all higher-order moments
$M_p(\mathbf{k})=\mathbb{E}[J_{\mathbf{k}}^p]$
of single-mode amplitudes are unchanged. To see this, it is easiest to use the equivalent formulation
of our theory in terms of the equation \eqref{Feq} for the free-energy functional $F[\lambda].$ The latter
is the generating functional for all $p$th-order cumulants:
\begin{equation}
C_p(\bm{\xi}_1,...,\bm{\xi}_p)
=
\frac{\delta^p F[\lambda]}{\delta\lambda(\bm{\xi}_1)\cdots\delta\lambda(\bm{\xi}_p)}
\Big\rvert_{\lambda=0},
\end{equation}
so that it is straightforward to obtain dynamical equations for all such cumulants
by taking functional derivatives of the equation \eqref{Feq} for $F.$ In this setting, the
single-mode statistics studied by \cite{lvov2004noisy} correspond to singular terms
$C_p(\bm{\xi})$ with all wave-numbers coinciding:
\begin{equation}
C_p(\bm{\xi}_1,...,\bm{\xi}_p)=C_p(\bm{\xi}_1)
\prod_{j>1} \delta^d(\bm{\xi}_j-\bm{\xi}_1) + \bar{C}_p(\bm{\xi}_1,...,\bm{\xi}_p)
\end{equation}
where the part $\bar{C}_p$ corresponds to contributions which are smooth or less singular (only subsets
of momenta coinciding). Without giving details, we note that the
single-mode cumulant equation derived from our theory
\begin{equation}
\dot{C}_p(\bm{\xi})=-p\gamma[n](\bm{\xi}) C_p(\bm{\xi})
+2(2p-3)\eta[n](\bm{\xi})C_{p-1}(\bm{\xi})
,\quad p\geq 2
\end{equation}
is equivalent to the moment equation derived in \cite{lvov2004noisy}; see Eq. (11) there.
}

\red{Our theory thus recovers the predictions from prior work, but it also predicts
new effects due to the statistical correlations between distinct wave-modes.
For example, for structure functions of the underlying wave field,
$S_p(\bm{r})={\mathbb E}[\lvert\Psi(\bm{x+r})-\Psi(\bm{x})\rvert^p],$
one obtains asymptotic relations in the kinetic limit, such as
\begin{eqnarray*}
S_4(\bm{r}) &\sim & 2\left[S_2(\bm{r})\right]^2
             +\left(\frac{2\pi}{L}\right)^d \int \mathrm{d}^d\bm{\xi} \,\,
                    \left[C_2(\bm{\xi})-2n^2(\bm{\xi})\right]
                    \lvert e^{i\bm{\xi}\cdot\bm{r}}-1\rvert^4 \cr
             && +2\left(\frac{2\pi}{L}\right)^d \int d^d\bm{\xi} \int d^d\bm{\xi}' \,\,
                     \bar{C}_2(\bm{\xi},\bm{\xi}')
                   \lvert e^{i\bm{\xi}\cdot\bm{r}}-1\rvert^2 \lvert e^{i\bm{\xi}'\cdot\bm{r}}-1\rvert^2 + o(L^{-d})
\end{eqnarray*}
where the first two terms on the righthand side have been previously discussed (see \cite{Nazarenko2011}, Eq.(5.32)),
but the  final term due to mode-correlations is new. Such effects due to correlated noise} play potentially an important role
in presence of forcing and dissipation where the time-reversal symmetry is broken.
\black{
Moreover, even though both the large deviation dynamics
have the same quasipotential at equilibrium, one notices that the time-inversal symmetry relation \eqref{eq:LD_detailed_balance} that must hold at equilibrium is broken for the weak-noise mean-field dynamics prescribed by \eqref{eq:H_mean-field}. This confirms that the weak-noise dynamics of the empirical spectrum described by the mean-field Hamiltonian \eqref{eq:H_mean-field} is not physically relevant.
}


\section{Out-of-equilibrium: adding forcing and dissipation}
\label{sec:out-of-eq}

\black{To allow for turbulent cascade solutions of the kinetic equation, driving and damping terms must be
added to the dynamics. The simplest approach is to add a weak, slowly varying linear term to the microscopic
wave equation \eqref{eq:Ak_evol_eq}, of the form \cite{EyinkShi2012}
\begin{align}
\frac{\text{d}A_{\bm{k}}}{\text{d}t} & =
\cdots + \frac{1}{2} \epsilon^2\Gamma(\bm{k},\epsilon^2 t) A_{\bm{k}}   \label{eq:Ak_evol_outofeq}\, .
\end{align}
\black{Here, for any wavenumber $\bm{k},$ a value $\Gamma(\bm{k}, \tau)>0$ corresponds to parametric forcing (typically at
small wavenumbers) and $\Gamma(\bm{k},\tau)<0$ corresponds to damping (typically at large wavenumbers). There
generally exists an intermediate range, called inertial range, for which $\Gamma(\bm{k}, \tau)=0$.}
Then all of the derivations in this paper carry through, with the large-deviations Hamiltonian
acquiring a new explicitly time-dependent term
\begin{equation} H[n,\lambda,\tau]=\cdots + \int d^d \xi \, \lambda(\bm{\xi}) \Gamma(\bm{\xi},\tau) n(\bm{\xi}).
\end{equation}
In that case, the Onsager-Machlup Lagrangian \eqref{OM-lag} remains the same except for the replacement
$\dot{n}(\bm{\xi},\tau)-\mathcal{C}[n](\bm{\xi}) \rightarrow \dot{n}(\bm{\xi},\tau)-\mathcal{C}[n](\bm{\xi})-\Gamma(\bm{\xi},\tau)n(\bm{\xi},\tau), $
so that the most-probable behavior corresponds to the solution of the modified kinetic equation
\begin{equation}
  \dot{n}(\bm{\xi},\tau)=\mathcal{C}[n](\bm{\xi}) + \Gamma(\bm{\xi},\tau)n(\bm{\xi},\tau).
  \label{turb-Keq}
\end{equation}
\black{The out-of-equilibrium term $\Gamma(\bm{\xi},\tau)n(\bm{\xi},\tau)$ allows turbulent Kolmogorov-Zakharov (KZ) solutions
at a range of intermediate wavenumbers (inertial range)\cite{Zakharov1992}, with energy flux (cascade) across scales.}
}

\black{An out-of-equilibrium action can thus be easily derived, at least at a formal level.
However, we know that KZ spectra are generally not valid \red{either} at high or low wavenumbers,
where the condition of time-scale separation \eqref{eq:RP_condition} breaks down;
\red{see \cite{biven2001breakdown,newell2001wave,connaughton2003dimensional} and
further discussion in} \cite{newell2011wave}. \red{The non-uniformity in wavenumber
of the condition \eqref{eq:RP_condition} imposes important restrictions in non-equilibrium settings,
since wave kinetics often maintains validity only in a certain finite range of wave-numbers and the modes
in that range are then essentially coupled to modes outside that range where a non-kinetic
description prevails, e.g. the weakly nonlinear inverse cascade in surface-gravity wave turbulence
may terminate in nonlinear dissipative structures such as sharp-crested waves
\cite{falcon2020saturation}. In numerical studies such non-uniformity may be accommodated by
choosing forcing and/or damping so that the kinetic description is valid over the entire simulated
wavenumber range, e.g. see \cite{vladimirova2021turbulence}. For rigorous mathematical studies
of non-equilibrium turbulent regimes, however, this non-uniform validity of wave kinetics poses a significant
difficulty.}  The regime of validity of the large deviation theory presented here thus deserves a more careful
analysis, that we defer to a future work.

Out of equilibrium, the detailed-balance relation \eqref{eq:LD_detailed_balance} no longer holds. Therefore, the fluctuation-dissipation relation
is broken and the noise correlation term (quadratic term in $\lambda$
in the Hamiltonian (\ref{eq:H_final_result})) generally affects the stationary distribution. In particular,
one expects the out-of-equilibrium quasipotential to depends explicitly on this
noise correlation term. Although a general explicit expression for an out-of-equilibrium quasipotential does not
seem attainable, looking for a perturbative solution with respect to the forcing strength \cite{bouchet2016perturbative} may provide a lead.

To conclude, one should emphasize that the use of the large deviation Hamiltonian \eqref{eq:H_final_result} is not restricted to equilibrium.
While being aware of possible restrictions on its range of validity, one can in principle use it within the inertial range to estimate the probability
of rare fluctuations from non-equilibrium spectra.} \red{Of course, the theory should describe {\it fortiori} the small fluctuations
$\delta n(\bm{\xi}):=\hat{n}(\bm{\xi})-n(\bm{\xi})$ of the empirical spectrum around the solution of the kinetic equation \eqref{turb-Keq},
which have typical magnitude $\delta n \sim 1/L^{d/2}.$ The predictions at this level (central limit theorem) are those of
a linear Langevin model with additive noise:
\begin{eqnarray*}
\frac{\mathrm{d}}{\mathrm{d}\tau}\delta n(\bm{\xi})&=&{\mathcal L}\delta n(\bm{\xi},\tau) +
   \Gamma(\bm{\xi},\tau)\delta n(\bm{\xi},\tau) \cr
   && \hspace{30pt} + \left(\frac{2\pi}{L}\right)^{d/2}
    \int \mathrm{d}^d\bm{\xi}' \,\, (2Q)^{1/2}[n](\bm{\xi},\bm{\xi}') \mathrm{d}W(\bm{\xi}',\tau),  \label{CLT-Langevin}
 \end{eqnarray*}
where  ${\mathcal L}$ is the collision operator linearized around the solution $n$:
\begin{eqnarray*}
&& {\mathcal L}\delta n(\bm{\xi}) := 36\pi\sum_{\substack{\vec{\sigma}\\\sigma_{1}=-1}} \int \mathrm{d}^d \bm{\xi}_2 \mathrm{d}^d\bm{\xi}_3\,
\lvert V^{\vec{\sigma}}_{\vec{\bm{\xi}}}\rvert^2
\delta(\vec{\sigma}\cdot\omega(\vec{\bm{\xi}}))\delta^d(\vec{\sigma}\cdot\vec{\bm{\xi}})\cr
&&\times \Big[ \delta n(\bm{\xi}_2)n(\bm{\xi}_3)+n(\bm{\xi}_2)\delta n(\bm{\xi}_3)
-\left(\sigma_2 \delta n(\bm{\xi})n(\bm{\xi}_3)+ \sigma_2 n(\bm{\xi})\delta n(\bm{\xi}_3)+(2\leftrightarrow 3)\right)\Big]
\end{eqnarray*}
}
\red{All of these possibilities} should be investigated in future work.

\section{Conclusion}\label{sec:Conclusion}

The kinetic equation describing the evolution of the spectrum in the
kinetic limit of weak wave turbulence is well established \black{and recently
rigorously derived in some cases.} However, to our knowledge, the statistics
of the empirical spectrum that have been studied so far in the kinetic
limit have not acknowledged the large deviation scaling that naturally
appears. We fill this gap in this paper and propose (based on previous
computations of the moment generating function of the mode amplitudes)
a large deviation Hamiltonian that quantifies the path probability
of the empirical spectral density in the kinetic limit (under the
RP approximation). \red{The exact form of the Hamiltonian is given for $3$-wave interacting systems,
but we expect that our result can be straightforwardly generalized to higher-order nonlinearities;
see section \ref{higher}.}
The Hamiltonian is associated with a generalized weak-noise Langevin dynamics for the empirical spectral density.

Fundamental properties of
the Hamiltonian \eqref{eq:H_final_result} have been checked. On the one hand, the two fundamental conservation
laws (energy and momentum) that exist for 3-wave interacting systems
in the kinetic limit are shown to be satisfied at the large deviation level.
On the other hand, the quasipotential with which the large deviation
dynamics satisfies a global detailed balance relation is shown to
be the quasipotential derived from the equilibrium microcanonical
distribution for fixed energy and momentum.

We have also compared the Hamiltonian \eqref{eq:H_final_result} to a mean-field Hamiltonian \eqref{eq:H_mean-field}
derived from \black{a microscopic dynamics inspired from the \emph{Random Phase and Amplitude} (RPA) approximation,
according to which the spectrum follows the very same kinetic equation \eqref{eq:kinetic_eq}.}
Although both Hamiltonians appear to be
very close in their expressions, the presence of coupling between
different wavevectors in the noise correlations breaks the pure mean-field
interaction \black{and allows one to recover the expected time-reversal symmetry that must be satisfied in equilibrium}.

\black{Finally, we have sketched the derivation of the large deviation Hamiltonian in presence of forcing and dissipation terms of the form \eqref{eq:Ak_evol_outofeq}.}
\black{One interesting physical application of this theory is to predict the transition time to a new state
when the Kolmogorov-Zakharov cascade solution is unstable. As discussed in \cite{Zakharov1992},
Ch. 4, the homogeneous and isotropic cascade solution may be unstable to small perturbations that
break such symmetries. This phenomenon has recently been studied numerically in weak turbulence
of capillary waves on shallow water \cite{vladimirova2021turbulence}, where it was found
with \black{anisotropic but reflection-symmetric forcing} that the isotropic cascade solution is
unstable and spontaneously breaks the reflection symmetry. A similar phenomena was observed long
ago in the 4-wave Majda-McLaughlin-Tabak (MMT) model, where the Kolmogorov-Zakharov wave turbulence
solution was found to be unstable to spatially-inhomogeneous perturbations and the resulting solution
spontaneously broke the translation-symmetry of the dynamics \cite{newell2012spontaneous}.
The typical spontaneous random fluctuations described by our theory are very tiny, of order $1/L^{d/2},$
or the inverse square-root of the flow volume. In the limit for which
wave-kinetics is valid (\red{$\epsilon \sim L^{\kappa/2}$, $0<\kappa \leqslant 1$}) these spontaneous fluctuations are much smaller \red{(for $d>1$)}
than the next-order corrections in the weak nonlinearity, which are expected to be of relative order $\epsilon$
\cite{benney1967sequential,erofeev1989kinetics}. However, these larger corrections from weak nonlinearity
will preserve all of the symmetries of the base solution and of the underlying wave dynamics. Thus, the spontaneous
random fluctuations described by our theory are the most significant intrinsic source  of symmetry-breaking
perturbations, which can seed a transition even when all environmental sources of perturbation are negligible.
Note that for application to the MMT model our theory would need to be generalized to 4-wave interactions
\cite{chibbaro2017wave} and to spatially-inhomogeneous wave kinetics \cite{ampatzoglou2021derivation}, \black{which we aim at investigating in the near future}.}

\black{Whenever a situation of multistability exists, path large deviation results are extremely important as they
are a key step in determining the transition rates and transition paths between different attractors. For the
wave turbulence kinetic equation, such multistability would be present if two different stable stationary solutions
should exist at the same time. This is possible in principle, although this has not been observed for the wave
turbulence kinetic theory, as far as we know. Could the symmetry-broken solutions discussed in previous works
\cite{Zakharov1992,vladimirova2021turbulence,newell2012spontaneous} be associated to multistability? This
seems a very natural hypothesis that has not been studied so far. Such a case of bistability would be an example
where very tiny stochastic fluctuations are crucial. The next-order corrections to the deterministic part (the kinetic
equation) in the weak nonlinearity, are expected to be of relative order $\epsilon$. In such a situation, the instantons
which describe the most probable paths from one attractor to another, or the action which is important for computing
the transition rates, would be given at leading order by a balance between the principal deterministic part and
the very small noise. Then even if the random part, of order $1/L^{d/2}$, is much smaller that the first deterministic
correction of order $\epsilon$, it would be responsible for the quantitative description of the transitions at leading order.}

\backmatter

\bmhead{Acknowledgments}

This work was supported by the Simons Foundation through the \black{Collaboration Grant 651463 ``Wave Turbulence'' (F.B. and J.G.)
and the Targeted Grant in MPS 663054 ``Revisiting the Turbulence Problem Using Statistical Mechanics'' (F.B. and G.E.)}. We thank Yu Deng, Zaher Hani,
Sergey Nazarenko, Alan Newell, Laure Saint-Raymond, and Malo Tarpin for interesting discussions on the topic of weak turbulence.

\bmhead{Data availability statement}
Data sharing not applicable to this article as no datasets were generated or analysed during the current study.

\bmhead{Conflicts of interest}
The authors have no relevant financial or non-financial interests to disclose.

\begin{appendices}

\section{Perturbative expansion of the moment generating function $Z_{L,\epsilon}$ \eqref{eq:def_mgf} }\label{sec:Appendix-B}

Our goal is to compute the moment generating function $Z_{L,\epsilon}$
at an intermediate time $\Delta t$ (which satisfies $1\ll\Delta t\ll\epsilon^{-2}$ as prescribed by \eqref{eq:kinetic_ineq}), from time $t=0$, conditioned on $\hat{n}(\bm{\xi},0)=n(\bm{\xi})$, or equivalently, $\{ \left\lvert b_{\bm{k}}\right\rvert^{2} \}$. We recall that $\mathbb{E}_n$ refers to the average with respect to the uniform measure over the phases (RP).

To do so, we will look for a perturbation expansion in $\epsilon$
of the dynamics (\ref{eq:bk_evol_eq}).

\subsection{Perturbative expansion of the modes~\eqref{eq:bk_evol_eq}}

In order to anticipate the continuous limit $L\to\infty$, we will
consider a slightly modified but equivalent evolution equation (\ref{eq:bk_evol_eq}).
The Kronecker-$\delta$ will be replaced by $\text{\ensuremath{\left(\tfrac{2\pi}{L}\right)^{d}}}\chi_{L}^{d}$,
with $\chi_{L}^{d}$ the normalized characteristic function of the
set $\left[-\tfrac{\pi}{L},\tfrac{\pi}{L}\right]^{d}$, defined as
\[
\chi_{L}^{d}(\bm{x})=\begin{cases}
\left(\tfrac{L}{2\pi}\right)^{d} & \text{if }\bm{x}\in\left[-\tfrac{\pi}{L},\tfrac{\pi}{L}\right]^{d}\\
0 & \text{otherwise}
\end{cases}.
\]
$\chi_{L}^{d}(\bm{x})$ is then a precursor of the Dirac-$\delta$
in $d$-dimension: $\chi_{L}^{d}\underset{L\to\infty}{\rightarrow}\delta^{d}$.
Therefore, Eq. (\ref{eq:bk_evol_eq}) is replaced by
\begin{equation}
i\frac{\text{d}b_{\bm{k}_{1}}}{\text{d}t} =3\epsilon\left(\frac{2\pi}{L}\right)^{3d/2}\sum_{\substack{\vec{\sigma}_{123}\\
\sigma_{1}=-1
}
}\sum_{\bm{k}_{2},\bm{k}_{3}}V_{\vec{\bm{k}}_{123}}^{\vec{\sigma}_{123}}b_{\bm{k}_{2}}^{\sigma_{2}}b_{\bm{k}_{3}}^{\sigma_{3}}{\rm e}^{-i\left(\vec{\sigma}_{123}\cdot\vec{\omega}_{123}\right)t}\chi_{L}^{d}(\vec{\sigma}_{123}\cdot\vec{\bm{k}}_{123})   \label{eq:bk_evol_eq_continuous_L}
\end{equation}
Since a detailed book keeping of the indices will be important
in the sequel, we have introduced a new notation to refer to a triad: $\vec{x}_{lmn}=(x_{l},x_{m},x_{n})$.
We will note with upper indices the components that take a minus sign.
For instance $(x_{1},-x_{2},x_{3})$ will be denoted $\vec{x}_{13}^{2}$.
Indices will be always labelled in ascending order to avoid confusion
on their position within the triplet $(x_{l},x_{m},x_{n})$.

We are now ready to perform the perturbation expansion for the dynamics
(\ref{eq:bk_evol_eq_continuous_L}). We look for a solution $b_{\bm{k}}(t)$
for $t\geqslant0$ as a perturbation expansion in $\epsilon$:
\begin{equation}
b_{\bm{k}}(t)=b_{\bm{k}}^{(0)}(t)+\epsilon b_{\bm{k}}^{(1)}(t)+\epsilon^{2}b_{\bm{k}}^{(2)}(t)+\mathcal{O}\left(\epsilon^{3}\right)\quad.\label{eq:pert_exp_bk}
\end{equation}
with $b_{k}(0)=b_{k}^{(0)}(0)$.

Integrating from $t=0$ to $t=\Delta t$, one gets the following hierarchy:
\begin{align}
& b_{\bm{k}_{1}}^{(0)}(\Delta t)  = b_{\bm{k}}^{(0)}(0)\label{eq:order_0_bk_expansion}\\
& b_{\bm{k}_{1}}^{(1)\sigma_{1}}(\Delta t)  \label{eq:order_1_bk_expansion} \\
& \; =-3i\left(\frac{2\pi}{L}\right)^{3d/2}  \!\!\! \sigma_{1}\sum_{\sigma_{2},\sigma_{3}}\sum_{\bm{k}_{2},\bm{k}_{3}}V_{\vec{\bm{k}}_{123}}^{\vec{\sigma}_{23}^{1}}b_{\bm{k}_{2}}^{(0)\sigma_{2}}b_{\bm{k}_{3}}^{(0)\sigma_{3}}\Delta_{\Delta t}\left(-\vec{\sigma}_{23}^{1}\cdot\vec{\omega}_{23}^{1}\right)\chi_{L}^{d}(\vec{\sigma}_{23}^{1}\cdot\vec{\bm{k}}_{23}^{1}) \nonumber \\
& b_{\bm{k}_{1}}^{(2)}(\Delta t)  \label{eq:order_2_bk_exp_intermediate} \\
 &\;  =-6i\left(\frac{2\pi}{L}\right)^{3d/2} \!\! \sum_{\substack{\vec{\sigma} \\ \sigma_{1}=-1}}
  \sum_{\bm{k}_{2},\bm{k}_{3}}V_{\vec{\bm{k}}_{123}}^{\vec{\sigma}_{123}}b_{\bm{k}_{2}}^{(0)\sigma_{2}} \nonumber \\
  & \hspace{4cm} \times \left(\int_{0}^{\Delta t}b_{\bm{k}_{3}}^{(1)\sigma_{3}}(t){\rm e}^{-i(\vec{\sigma}_{123}\cdot\vec{\omega}_{123})t}{\rm d}t\right)\chi_{L}^{d}(\vec{\sigma}_{123}\cdot\vec{\bm{k}}_{123}) \nonumber
\end{align}
with
\[
\Delta_{T}(x)=\int_{0}^{T} \mathrm{e}^{ixt}\mathrm{d}t 
\]

Injecting (\ref{eq:order_1_bk_expansion}) into (\ref{eq:order_2_bk_exp_intermediate}) yields
\begin{align}
& b_{\bm{k}_{1}}^{(2)}(\Delta t)  =18\left(\frac{2\pi}{L}\right)^{3d} \label{eq:order_2_bk_expansion} \\
& \qquad  \times \!\!\sum_{\substack{\sigma_{2},\sigma_{3},\sigma_{4},\sigma_{5} \\
\sigma_{1}=-1
}
}
\!\!\!\sigma_{1}\sigma_{3}\!\!\!\sum_{\bm{k}_{2},\bm{k}_{3},\bm{k}_{4},\bm{k}_{5}}\left\{ V_{\vec{\bm{k}}_{123}}^{\vec{\sigma}_{123}}V_{\vec{\bm{k}}_{345}}^{\vec{\sigma}_{45}^{3}}\chi_{L}^{d}\left(\vec{\sigma}_{123}\cdot\vec{\bm{k}}_{123}\right)\chi_{L}^{d}\left(\vec{\sigma}_{45}^{3}\cdot\vec{\bm{k}}_{345}\right)\right. \nonumber \\
 & \qquad\qquad\qquad\left.\times\left(b_{\bm{k}_{2}}^{(0)\sigma_{2}}b_{\bm{k}_{4}}^{(0)\sigma_{4}}b_{\bm{k}_{5}}^{(0)\sigma_{5}}\right)\tilde{E}_{\Delta t}\left(-\vec{\sigma}_{123}\cdot\vec{\omega}_{123};-\vec{\sigma}_{45}^{3}\cdot\vec{\omega}_{345}\right)\right\} \nonumber
\end{align}

\begin{equation}
\tilde{E}_{T}(x,y)  =\int_{0}^{T}\mathrm{e}^{ixt}\Delta_{t}(y) \mathrm{d}t =-i\frac{\Delta_{T}(y+x)-\Delta_{T}(x)}{y}.
\end{equation}

\subsection{Computation of the moment generating function $Z_{L,\epsilon}$ (\ref{eq:def_mgf})}

From the definition of the moment generating function $Z_{L,\epsilon}$
\eqref{eq:def_mgf}, using \eqref{eq:def_empirical_spectrum}, one
obtains
\begin{equation}
Z_{L,\epsilon}  =\mathbb{E}_{n}\left[{\rm e}^{\sum_{\bm{k}}\lambda(\bm{k})\left(\left\lvert b_{\bm{k}}(\Delta t)\right\rvert^{2}-\left\lvert b_{\bm{k}}(0)\right\rvert^{2}\right)}\right]\, .
\end{equation}

From Eq. \eqref{eq:pert_exp_bk}, one obtains
\begin{align*}
\left\lvert b_{\bm{k}}(\Delta t)\right\rvert^{2} & =\left\lvert b_{\bm{k}}^{(0)}\right\rvert^{2}+\epsilon\left(b_{\bm{k}}^{(1)}b_{\bm{k}}^{(0)\ast}+b_{\bm{k}}^{(1)\ast}b_{\bm{k}}^{(0)}\right) \\
&  \hspace{3cm} +\epsilon^{2}\left(b_{\bm{k}}^{(2)}b_{\bm{k}}^{(0)\ast}+b_{\bm{k}}^{(2)\ast}b_{\bm{k}}^{(0)}+b_{\bm{k}}^{(1)}b_{\bm{k}}^{(1)\ast}\right)
+\mathcal{O}\left(\epsilon^{3}\right)\,.
\end{align*}
Hence, the expansion of $Z_{L,\epsilon}$ with respect to $\epsilon$
reads as
\begin{align}
Z_{L,\epsilon} & =\mathbb{E}_{n}\left[1+\epsilon G_{1}+\epsilon^{2}\left(G_{2}+\frac{1}{2}G_{1}^{2}\right)+\mathcal{O}\left(\epsilon^{3}\right)\right]\label{eq:mgf_Z_function_G1_G2}
\end{align}
with
\begin{align*}
G_{1} & =\sum_{\bm{k}}\lambda(\bm{k})\left(b_{\bm{k}}^{(1)}b_{\bm{k}}^{(0)\ast}+b_{\bm{k}}^{(1)\ast}b_{\bm{k}}^{(0)}\right)\\
G_{2} & =\sum_{\bm{k}}\lambda(\bm{k})\left(b_{\bm{k}}^{(2)}b_{\bm{k}}^{(0)\ast}+b_{\bm{k}}^{(2)\ast}b_{\bm{k}}^{(0)}+b_{\bm{k}}^{(1)}b_{\bm{k}}^{(1)\ast}\right)\,.
\end{align*}

\subsubsection{Computation of $\mathcal{O}\left(\epsilon^{1}\right)$ term}

From Eqs. \eqref{eq:order_0_bk_expansion} and \eqref{eq:order_1_bk_expansion},
$G_{1}$ reads as
\begin{align}
  G_{1} & = 3i\left(\frac{2\pi}{L}\right)^{3d/2}  \!\! \sum_{\vec{\sigma}_{123}}\sum_{\vec{\bm{k}}_{123}}\lambda(\bm{k}_{1})\sigma_{1} \label{eq:term_order_1_mgf_Z} \\
  & \quad \times \left[V_{\vec{\bm{k}}_{123}}^{\vec{\sigma}_{123}}b_{\bm{k}_{1}}^{(0)\sigma_{1}}b_{\bm{k}_{2}}^{(0)\sigma_{2}}b_{\bm{k}_{3}}^{(0)\sigma_{3}}\Delta_{\Delta t}\left(-\vec{\sigma}_{123}\cdot\vec{\omega}_{123}\right)\chi_{L}^{d}(\vec{\sigma}_{123}\cdot\vec{\bm{k}}_{123})\right]\;. \nonumber
\end{align}

However, since
\[
\mathbb{E}_{n}\left[b_{\bm{k}_{1}}^{(0)\sigma_{1}}b_{\bm{k}_{2}}^{(0)\sigma_{2}}b_{\bm{k}_{3}}^{(0)\sigma_{3}}\right]=0
\]
for any $\vec{\sigma}=(\sigma_{1},\sigma_{2},\sigma_{3})$, we deduce $\mathbb{E}_n\left[ G_1 \right]=0$, namely
that the $\mathcal{O}\left(\epsilon^{1}\right)$ term of $Z_{L,\epsilon}$
vanishes.

As a remark, one may notice that $\mathbb{E}_{n}\left[\prod_{i=1}^{p}b_{\bm{k}_{i}}^{(0)\sigma_{i}}\right]=0$
for any \emph{odd} integer $p$. We thus deduce that there is not
any correction of order $\epsilon^{p}$ ($p$ odd) in the moment generating
function $Z_{L,\epsilon}$.

\subsubsection{Computation of $\mathcal{O}\left(\epsilon^{2}\right)$ term }

\paragraph{Expressions of $G_{2}$ and $G_{1}^{2}$}

From Eqs. (\ref{eq:order_0_bk_expansion}), (\ref{eq:order_1_bk_expansion})
and (\ref{eq:order_2_bk_exp_intermediate}), $G_{2}$ reads as
\begin{align*}
  G_{2} & =\left(\frac{2\pi}{L}\right)^{3d}\sum_{\bm{k}_{1}}\lambda(\bm{k}_{1}) \\
        & \quad \times \Bigg\{ 18 \!\!\!\! \sum_{\sigma_{1},\sigma_{2},\sigma_{3},\sigma_{4},\sigma_{5}} \!\!\! \sigma_{1}\sigma_{3} \!\!\! \sum_{\bm{k}_{2},\bm{k}_{3},\bm{k}_{4},\bm{k}_{5}} \!\!\!  V_{\vec{\bm{k}}_{123}}^{\vec{\sigma}_{123}}V_{\vec{\bm{k}}_{345}}^{\vec{\sigma}_{45}^{3}}\chi_{L}^{d}(\vec{\sigma}_{123}\cdot\vec{\bm{k}}_{123})\chi_{L}^{d}(\vec{\sigma}_{45}^{3}\cdot\vec{\bm{k}}_{345}) \nonumber\\
 & \hspace{2.1cm} \times\left(b_{\bm{k}_{1}}^{(0)\sigma_{1}}b_{\bm{k}_{2}}^{(0)\sigma_{2}}b_{\bm{k}_{4}}^{(0)\sigma_{4}}b_{\bm{k}_{5}}^{(0)\sigma_{5}}\right)\tilde{E}_{\Delta t}\left(-\vec{\sigma}_{123}\cdot\vec{\omega}_{123};-\vec{\sigma}_{45}^{3}\cdot\vec{\omega}_{345}\right)\\
 & \hspace{0.8cm} +\frac{9}{2}\sum_{\sigma_{1},\sigma_{2},\sigma_{3},\sigma_{4},\sigma_{5}} \!\!\! \sigma_{1}^{2} \!\! \sum_{\bm{k}_{2},\bm{k}_{3},\bm{k}_{4},\bm{k}_{5}}V_{\vec{\bm{k}}_{123}}^{\vec{\sigma}_{23}^{1}}V_{\vec{\bm{k}}_{145}}^{\vec{\sigma}_{1}^{45}}\left(b_{\bm{k}_{2}}^{(0)\sigma_{2}}b_{\bm{k}_{3}}^{(0)\sigma_{3}}b_{\bm{k}_{4}}^{(0)-\sigma_{4}}b_{\bm{k}_{5}}^{(0)-\sigma_{5}}\right)\\
 & \hspace{2.1cm} \times\Delta_{\Delta t}\left(-\vec{\sigma}_{23}^{1}\cdot\vec{\omega}_{123}\right)\Delta_{\Delta t}^{\ast}\left(\vec{\sigma}_{1}^{45}\cdot\vec{\omega}_{145}\right)\chi_{L}^{d}(\vec{\sigma}_{23}^{1}\cdot\vec{\bm{k}}_{123})\chi_{L}^{d}(\vec{\sigma}_{1}^{45}\cdot\vec{\bm{k}}_{145}) \Bigg\}
\end{align*}
Furthermore, one gets from Eq. \eqref{eq:term_order_1_mgf_Z}:
\begin{align*}
  G_{1}^{2} & =-9\left(\frac{2\pi}{L}\right)^{3d} \!\! \sum_{\vec{\sigma}_{123},\vec{\sigma}_{456}}\sum_{\vec{\bm{k}}_{123},\vec{\bm{k}}_{456}}\sigma_{1}\lambda(\bm{k}_{1})\sigma_{4}\lambda(\bm{k}_{4}) \\
            & \hspace{1.5cm} \times \left[V_{\vec{\bm{k}}_{123}}^{\vec{\sigma}_{123}}b_{\bm{k}_{1}}^{(0)\sigma_{1}}b_{\bm{k}_{2}}^{(0)\sigma_{2}}b_{\bm{k}_{3}}^{(0)\sigma_{3}}\Delta_{\Delta t}\left(\vec{\sigma}_{123}\cdot\vec{\omega}_{123}\right)\chi_{L}^{d}(\vec{\sigma}_{123}\cdot\vec{\bm{k}}_{123})\right]\\
 & \hspace{1.5cm} \times\left[V_{\vec{\bm{k}}_{456}}^{\vec{\sigma}_{456}}b_{\bm{k}_{4}}^{(0)\sigma_{4}}b_{\bm{k}_{5}}^{(0)\sigma_{5}}b_{\bm{k}_{6}}^{(0)\sigma_{6}}\Delta_{\Delta t}\left(\vec{\sigma}_{456}\cdot\vec{\omega}_{456}\right)\chi_{L}^{d}(\vec{\sigma}_{456}\cdot\vec{\bm{k}}_{456})\right]
\end{align*}

\paragraph{Random phase averages $\mathbb{E}_{n}\left[G_{2}\right]$ and $\mathbb{E}_{n}\left[G_{1}^{2}\right]$}
The contributions $G_{2}$ and $G_{1}^{2}$ contain respectively terms
of order $4$ and $6$ in $b_{\bm{k}}$, whose average over the uniform
phase distribution (RP) yields non zero contributions.
$\mathbb{E}_{n}\left[b_{\bm{k}_{1}}^{(0)\sigma_{1}}b_{\bm{k}_{2}}^{(0)\sigma_{2}}b_{\bm{k}_{3}}^{(0)\sigma_{3}}b_{\bm{k}_{4}}^{(0)\sigma_{4}}\right]$
and $\mathbb{E}_{n}\left[b_{\bm{k}_{1}}^{(0)\sigma_{1}}b_{\bm{k}_{2}}^{(0)\sigma_{2}}b_{\bm{k}_{3}}^{(0)\sigma_{3}}b_{\bm{k}_{4}}^{(0)\sigma_{4}}b_{\bm{k}_{5}}^{(0)\sigma_{5}}b_{\bm{k}_{6}}^{(0)\sigma_{6}}\right]$
will contain non zero contributions as long as the number of $b$
matches the number of $b^{\ast}$ in order to form non oscillating
terms.
\begin{itemize}
\item The first term in $G_{2}$ is $\mathbb{E}_{n}\left[b_{\bm{k}_{1}}^{(0)\sigma_{1}}b_{\bm{k}_{2}}^{(0)\sigma_{2}}b_{\bm{k}_{4}}^{(0)\sigma_{4}}b_{\bm{k}_{5}}^{(0)\sigma_{5}}\right]$.
The latter can be explicitly computed under the RP distribution. It
contains three non-vanishing terms only:
\begin{align*}
 & \mathbb{E}_{n}\left[b_{\bm{k}_{1}}^{(0)\sigma_{1}}b_{\bm{k}_{2}}^{(0)\sigma_{2}}b_{\bm{k}_{4}}^{(0)\sigma_{4}}b_{\bm{k}_{5}}^{(0)\sigma_{5}}\right]\\
 & \quad=\delta_{\sigma_{1},-\sigma_{2}}\delta_{\sigma_{4},-\sigma_{5}}\delta_{\bm{k}_{1},\bm{k}_{2}}\delta_{\bm{k}_{4},\bm{k}_{5}}\left\lvert b_{\bm{k}_{1}}^{(0)}\right\rvert^{2}\left\lvert b_{\bm{k}_{4}}^{(0)}\right\rvert^{2}\\
 & \quad\quad+\left(\delta_{\sigma_{1},-\sigma_{4}}\delta_{\sigma_{2},-\sigma_{5}}\delta_{\bm{k}_{1},\bm{k}_{4}}\delta_{\bm{k}_{2},\bm{k}_{5}}+\delta_{\sigma_{1},-\sigma_{5}}\delta_{\sigma_{2},-\sigma_{4}}\delta_{\bm{k}_{1},\bm{k}_{5}}\delta_{\bm{k}_{2},\bm{k}_{4}}\right)\left\lvert b_{\bm{k}_{1}}^{(0)}\right\rvert^{2}\left\lvert b_{\bm{k}_{2}}^{(0)}\right\rvert^{2}
\end{align*}
\item The second term in $G_{2}$ is $\mathbb{E}_{n}\left[b_{\bm{k}_{2}}^{(0)\sigma_{2}}b_{\bm{k}_{3}}^{(0)\sigma_{3}}b_{\bm{k}_{4}}^{(0)-\sigma_{4}}b_{\bm{k}_{5}}^{(0)-\sigma_{5}}\right]$
and reads as
\begin{align*}
 & \mathbb{E}_{n}\left[b_{\bm{k}_{2}}^{(0)\sigma_{2}}b_{\bm{k}_{3}}^{(0)\sigma_{3}}b_{\bm{k}_{4}}^{(0)-\sigma_{4}}b_{\bm{k}_{5}}^{(0)-\sigma_{5}}\right]\\
 & \quad=\delta_{\sigma_{2},-\sigma_{3}}\delta_{\sigma_{4},-\sigma_{5}}\delta_{\bm{k}_{2},\bm{k}_{3}}\delta_{\bm{k}_{4},\bm{k}_{5}}\left\lvert b_{\bm{k}_{2}}^{(0)}\right\rvert^{2}\left\lvert b_{\bm{k}_{4}}^{(0)}\right\rvert^{2}\\
 & \quad\quad+\left(\delta_{\sigma_{2},\sigma_{4}}\delta_{\sigma_{3},\sigma_{5}}\delta_{\bm{k}_{2},\bm{k}_{4}}\delta_{\bm{k}_{3},\bm{k}_{5}}+\delta_{\sigma_{2},\sigma_{5}}\delta_{\sigma_{3},\sigma_{4}}\delta_{\bm{k}_{2},\bm{k}_{5}}\delta_{\bm{k}_{3},\bm{k}_{4}}\right)\left\lvert b_{\bm{k}_{2}}^{(0)}\right\rvert^{2}\left\lvert b_{\bm{k}_{3}}^{(0)}\right\rvert^{2}
\end{align*}
\item Finally, the only term in $G_{1}^{2}$ is $\mathbb{E}_{n}\left[b_{\bm{k}_{1}}^{(0)\sigma_{1}}b_{\bm{k}_{2}}^{(0)\sigma_{2}}b_{\bm{k}_{3}}^{(0)\sigma_{3}}b_{\bm{k}_{4}}^{(0)\sigma_{4}}b_{\bm{k}_{5}}^{(0)\sigma_{5}}b_{\bm{k}_{6}}^{(0)\sigma_{6}}\right]$.
It contains 15 non-vanishing terms:
\begin{align*}
 & \mathbb{E}_{n}\left[b_{\bm{k}_{1}}^{(0)\sigma_{1}}b_{\bm{k}_{2}}^{(0)\sigma_{2}}b_{\bm{k}_{3}}^{(0)\sigma_{3}}b_{\bm{k}_{4}}^{(0)\sigma_{4}}b_{\bm{k}_{5}}^{(0)\sigma_{5}}b_{\bm{k}_{6}}^{(0)\sigma_{6}}\right]\\
  & \quad = \delta_{\sigma_{1},-\sigma_{2}}\delta_{\bm{k}_{1},\bm{k}_{2}}\left\lvert b_{\bm{k}_{1}}^{(0)}\right\rvert^{2}
    \left( \delta_{\sigma_{3},-\sigma_{4}}\delta_{\sigma_{5},-\sigma_{6}}\delta_{\bm{k}_{3},\bm{k}_{4}}\delta_{\bm{k}_{5},\bm{k}_{6}}\left\lvert b_{\bm{k}_{3}}^{(0)}\right\rvert^{2}\left\lvert b_{\bm{k}_{5}}^{(0)}\right\rvert^{2} \right.\\
  & \hspace{4.2cm} + \delta_{\sigma_{3},-\sigma_{5}}\delta_{\sigma_{4},-\sigma_{6}}\delta_{\bm{k}_{3},\bm{k}_{5}}\delta_{\bm{k}_{4},\bm{k}_{6}}\left\lvert b_{\bm{k}_{3}}^{(0)}\right\rvert^{2}\left\lvert b_{\bm{k}_{4}}^{(0)}\right\rvert^{2}  \\
  & \hspace{4.2cm} \left. + \, \delta_{\sigma_{3},-\sigma_{6}}\delta_{\sigma_{4},-\sigma_{5}}\delta_{\bm{k}_{3},\bm{k}_{6}}\delta_{\bm{k}_{4},\bm{k}_{5}}\left\lvert b_{\bm{k}_{3}}^{(0)}\right\rvert^{2}\left\lvert b_{\bm{k}_{4}}^{(0)}\right\rvert^{2} \right)\\
  & \qquad +\delta_{\sigma_{1},-\sigma_{3}}\delta_{\bm{k}_{1},\bm{k}_{3}}\left\lvert b_{\bm{k}_{1}}^{(0)}\right\rvert^{2}
    \left(\delta_{\sigma_{2},-\sigma_{4}}\delta_{\sigma_{5},-\sigma_{6}}\delta_{\bm{k}_{2},\bm{k}_{4}}\delta_{\bm{k}_{5},\bm{k}_{6}}\left\lvert b_{\bm{k}_{2}}^{(0)}\right\rvert^{2}\left\lvert b_{\bm{k}_{5}}^{(0)}\right\rvert^{2} \right. \\
  & \hspace{4.4cm} + \delta_{\sigma_{2},-\sigma_{5}}\delta_{\sigma_{4},-\sigma_{6}}\delta_{\bm{k}_{2},\bm{k}_{5}}\delta_{\bm{k}_{4},\bm{k}_{6}}\left\lvert b_{\bm{k}_{2}}^{(0)}\right\rvert^{2}\left\lvert b_{\bm{k}_{4}}^{(0)}\right\rvert^{2} \\
  & \hspace{4.4cm} \left. + \delta_{\sigma_{2},-\sigma_{6}}\delta_{\sigma_{4},-\sigma_{5}}\delta_{\bm{k}_{2},\bm{k}_{6}}\delta_{\bm{k}_{4},\bm{k}_{5}}\left\lvert b_{\bm{k}_{2}}^{(0)}\right\rvert^{2}\left\lvert b_{\bm{k}_{4}}^{(0)}\right\rvert^{2}\right)\\
  & \qquad +\delta_{\sigma_{1},-\sigma_{4}}\delta_{\bm{k}_{1},\bm{k}_{4}}\left\lvert b_{\bm{k}_{1}}^{(0)}\right\rvert^{2}
    \left(\delta_{\sigma_{2},-\sigma_{3}}\delta_{\sigma_{5},-\sigma_{6}}\delta_{\bm{k}_{2},\bm{k}_{3}}\delta_{\bm{k}_{5},\bm{k}_{6}}\left\lvert b_{\bm{k}_{2}}^{(0)}\right\rvert^{2}\left\lvert b_{\bm{k}_{5}}^{(0)}\right\rvert^{2} \right. \\
  & \hspace{4.4cm} + \delta_{\sigma_{2},-\sigma_{5}}\delta_{\sigma_{3},-\sigma_{6}}\delta_{\bm{k}_{2},\bm{k}_{5}}\delta_{\bm{k}_{3},\bm{k}_{6}}\left\lvert b_{\bm{k}_{2}}^{(0)}\right\rvert^{2}\left\lvert b_{\bm{k}_{3}}^{(0)}\right\rvert^{2} \\
  & \hspace{4.4cm} \left. +\delta_{\sigma_{2},-\sigma_{6}}\delta_{\sigma_{3},-\sigma_{5}}\delta_{\bm{k}_{2},\bm{k}_{6}}\delta_{\bm{k}_{3},\bm{k}_{5}}\left\lvert b_{\bm{k}_{2}}^{(0)}\right\rvert^{2}\left\lvert b_{\bm{k}_{3}}^{(0)}\right\rvert^{2}\right)\\
  & \qquad + \delta_{\sigma_{1},-\sigma_{5}}\delta_{\bm{k}_{1},\bm{k}_{5}}\left\lvert b_{\bm{k}_{1}}^{(0)}\right\rvert^{2}
    \left(\delta_{\sigma_{2},-\sigma_{3}}\delta_{\sigma_{4},-\sigma_{6}}\delta_{\bm{k}_{2},\bm{k}_{3}}\delta_{\bm{k}_{4},\bm{k}_{6}}\left\lvert b_{\bm{k}_{2}}^{(0)}\right\rvert^{2}\left\lvert b_{\bm{k}_{4}}^{(0)}\right\rvert^{2} \right.\\
  & \hspace{4.4cm} + \delta_{\sigma_{2},-\sigma_{4}}\delta_{\sigma_{3},-\sigma_{6}}\delta_{\bm{k}_{2},\bm{k}_{4}}\delta_{\bm{k}_{3},\bm{k}_{6}}\left\lvert b_{\bm{k}_{2}}^{(0)}\right\rvert^{2}\left\lvert b_{\bm{k}_{3}}^{(0)}\right\rvert^{2} \\
  & \hspace{4.4cm} \left. + \delta_{\sigma_{2},-\sigma_{6}}\delta_{\sigma_{3},-\sigma_{4}}\delta_{\bm{k}_{2},\bm{k}_{6}}\delta_{\bm{k}_{3},\bm{k}_{4}}\left\lvert b_{\bm{k}_{2}}^{(0)}\right\rvert^{2}\left\lvert b_{\bm{k}_{3}}^{(0)}\right\rvert^{2}\right)\\
  & \qquad + \delta_{\sigma_{1},-\sigma_{6}}\delta_{\bm{k}_{1},\bm{k}_{6}}\left\lvert b_{\bm{k}_{1}}^{(0)}\right\rvert^{2}
    \left(\delta_{\sigma_{2},-\sigma_{3}}\delta_{\sigma_{4},-\sigma_{5}}\delta_{\bm{k}_{2},\bm{k}_{3}}\delta_{\bm{k}_{4},\bm{k}_{5}}\left\lvert b_{\bm{k}_{2}}^{(0)}\right\rvert^{2}\left\lvert b_{\bm{k}_{4}}^{(0)}\right\rvert^{2} \right. \\
  & \hspace{4.4cm} + \delta_{\sigma_{2},-\sigma_{4}}\delta_{\sigma_{3},-\sigma_{5}}\delta_{\bm{k}_{2},\bm{k}_{4}}\delta_{\bm{k}_{3},\bm{k}_{5}}\left\lvert b_{\bm{k}_{2}}^{(0)}\right\rvert^{2}\left\lvert b_{\bm{k}_{3}}^{(0)}\right\rvert^{2} \\
  & \hspace{4.4cm} \left. +\delta_{\sigma_{2},-\sigma_{5}}\delta_{\sigma_{3},-\sigma_{4}}\delta_{\bm{k}_{2},\bm{k}_{5}}\delta_{\bm{k}_{3},\bm{k}_{4}}\left\lvert b_{\bm{k}_{2}}^{(0)}\right\rvert^{2}\left\lvert b_{\bm{k}_{3}}^{(0)}\right\rvert^{2}\right)
\end{align*}
\end{itemize}

Gathering all the contributions of $G_2$, we obtain
\begin{align*}
& \mathbb{E}_{n}\left[G_{2}\right] =\left(\frac{2\pi}{L}\right)^{2d}\sum_{\bm{k}_{1}}\lambda(\bm{k}_{1})\\
  & \hspace{0.5cm} \times\left\{ 36\sum_{\vec{\sigma}_{123}}\sigma_{1}\sigma_{3}\sum_{\bm{k}_{2},\bm{k}_{3}}\left\lvert V_{\vec{\bm{k}}_{123}}^{\vec{\sigma}_{123}}\right\rvert^{2}\left\lvert b_{\bm{k}_{1}}^{(0)}\right\rvert^{2}\left\lvert b_{\bm{k}_{2}}^{(0)}\right\rvert^{2}\chi_{L}^{d}(\vec{\sigma}_{123}\cdot\vec{\bm{k}}_{123}) \right.\\
  & \hspace{4cm} \times \tilde{E}_{\Delta t}\left(-\vec{\sigma}_{123}\cdot\vec{\omega}_{123};\vec{\sigma}_{123}\cdot\vec{\omega}_{123}\right) \\
 & \hspace{1cm} \left. + 9 \, \sum_{\vec{\sigma}_{123}}\sigma_{1}^{2}\sum_{\bm{k}_{2},\bm{k}_{3}}\left\lvert V_{\vec{\bm{k}}_{123}}^{\vec{\sigma}_{123}}\right\rvert^{2}\left\lvert b_{\bm{k}_{2}}^{(0)}\right\rvert^{2}\left\lvert b_{\bm{k}_{3}}^{(0)}\right\rvert^{2}\left\lvert \Delta_{\Delta t}\left(\vec{\sigma}_{123}\cdot\vec{\omega}_{123}\right)\right\rvert^{2}\chi_{L}^{d}(\vec{\sigma}_{123}\cdot\vec{\bm{k}}_{123})\right\} \\
 & \hspace{0.5cm} +R(G_{2})
\end{align*}
 with $R(G_{2})$ the remaining term coming from the internal pairing
of indices within the triads:
\begin{align*}
  R(G_{2}) & =\left(\frac{2\pi}{L}\right)^{2d}18\sum_{\sigma_{1},\sigma_{3},\sigma_{4}}\sum_{\bm{k}_{1},\bm{k}_{3},\bm{k}_{4}}\sigma_{1}\sigma_{3}\lambda(\bm{k}_{1})V_{\bm{k}_{1},\bm{k}_{1},\bm{k}_{3}}^{\sigma_{1},-\sigma_{1},\sigma_{3}}V_{\bm{k}_{3},\bm{k}_{4},\bm{k}_{4}}^{-\sigma_{3},\sigma_{4},-\sigma_{4}} \\
           &  \hspace{2.5cm} \times  \left\lvert b_{\bm{k}_{1}}^{(0)}\right\rvert^{2}\left\lvert b_{\bm{k}_{4}}^{(0)}\right\rvert^{2} \tilde{E}_{\Delta t}\left(-\sigma_{3}\omega(\bm{k}_{3});\sigma_{3}\omega(\bm{k}_{3})\right)\chi_{L}^{d}(\sigma_{3}\bm{k}_{3})\\
           & \; +\left(\frac{2\pi}{L}\right)^{2d}\frac{9}{2}\sum_{\sigma_{1},\sigma_{2},\sigma_{4}}\sum_{\bm{k}_{1},\bm{k}_{2},\bm{k}_{4}}\sigma_{1}^{2}\lambda(\bm{k}_{1})V_{\bm{k}_{1}, \bm{k}_2, \bm{k}_2}^{-\sigma_1,\sigma_2,-\sigma_2} V_{\bm{k}_{1}, \bm{k}_4, \bm{k}_4}^{\sigma_1,\sigma_4,-\sigma_4} \\
             & \hspace{2.5cm} \times \left\lvert b_{\bm{k}_{2}}^{(0)}\right\rvert^{2}\left\lvert b_{\bm{k}_{4}}^{(0)}\right\rvert^{2} \Delta_{\Delta t}\left(-\sigma_{1}\omega(\bm{k}_{1})\right)\Delta_{\Delta t}^{\ast}\left(\sigma_{1}\omega(\bm{k}_{1})\right) \chi_{L}^{d}(\sigma_{1}\bm{k}_{1})
\end{align*}
Note that $R(G_{2})$ vanishes because $\lim_{\bm{k}_{1}\to0}V_{\bm{k}_{1},\bm{k}_{2},\bm{k}_{3}}^{\vec{\sigma}}=0$
for any $\vec{\sigma}$ and any $\bm{k}_{2},\bm{k}_{3}$.
We have used the equality $\left[\chi_{L}^{d}(\bm{x})\right]^{2}=\left(\tfrac{L}{2\pi}\right)^{d}\chi_{L}^{d}(\bm{x})$
to get the proper scaling in $L$.

Similar calculations yields, for $\mathbb{E}\left[G_{1}^{2}\right]$:
\begin{align*}
  \mathbb{E}_{n}\left[G_{1}^{2}\right] & =18\left(\frac{2\pi}{L}\right)^{2d}\sum_{\vec{\sigma}_{123}}\sum_{\vec{\bm{k}}_{123}}\left(\sigma_{1}\lambda(\bm{k}_{1})\right)^{2}\left\lvert V_{\vec{\bm{k}}_{123}}^{\vec{\sigma}_{123}}\right\rvert^{2}\left\lvert b_{\bm{k}_{1}}^{(0)}\right\rvert^{2}\left\lvert b_{\bm{k}_{2}}^{(0)}\right\rvert^{2}\left\lvert b_{\bm{k}_{3}}^{(0)}\right\rvert^{2} \\
                                       & \hspace{4cm} \times \left\lvert \Delta_{\Delta t}\left(\vec{\sigma}_{123}\cdot\vec{\omega}_{123}\right)\right\rvert^{2}\chi_{L}^{d}(\vec{\sigma}_{123}\cdot\vec{\bm{k}}_{123})\\
                                       & \; +36\left(\frac{2\pi}{L}\right)^{2d}\sum_{\vec{\sigma}_{123}}\sum_{\vec{\bm{k}}_{123}}\left(\sigma_{1}\lambda(\bm{k}_{1})\right)\left(\sigma_{2}\lambda(\bm{k}_{2})\right)\left\lvert V_{\vec{\bm{k}}_{123}}^{\vec{\sigma}_{123}}\right\rvert^{2}\left\lvert b_{\bm{k}_{1}}^{(0)}\right\rvert^{2}\left\lvert b_{\bm{k}_{2}}^{(0)}\right\rvert^{2}\left\lvert b_{\bm{k}_{3}}^{(0)}\right\rvert^{2} \\
                                       & \hspace{4cm} \times \left\lvert \Delta_{\Delta t}\left(\vec{\sigma}_{123}\cdot\vec{\omega}_{123}\right)\right\rvert^{2}\chi_{L}^{d}(\vec{\sigma}_{123}\cdot\vec{\bm{k}}_{123})\\
 & \; +R(G_{1}^{2})\,.
\end{align*}
We do not write explicitly the contribution $R(G_{1}^{2})$ (that
is slightly lengthy) which comes from the internal pairing of indices
within triads (\emph{i.e.} all the terms in $\mathbb{E}_{n}\left[b_{\bm{k}_{1}}^{(0)\sigma_{1}}b_{\bm{k}_{2}}^{(0)\sigma_{2}}b_{\bm{k}_{3}}^{(0)\sigma_{3}}b_{\bm{k}_{4}}^{(0)\sigma_{4}}b_{\bm{k}_{5}}^{(0)\sigma_{5}}b_{\bm{k}_{6}}^{(0)\sigma_{6}}\right]$
except those proportional to $\left\lvert b_{\bm{k}_{1}}^{(0)}\right\rvert^{2}\left\lvert b_{\bm{k}_{2}}^{(0)}\right\rvert^{2}\left\lvert b_{\bm{k}_{3}}^{(0)}\right\rvert^{2}$
). The remainder $R(G_{1}^{2})$ is also vanishing because
of the property $\lim_{\bm{k}_{1}\to0}V_{\bm{k}_{1},\bm{k}_{2},\bm{k}_{3}}^{\vec{\sigma}}=0$
for any $\vec{\sigma}$ and any $\bm{k}_{2},\bm{k}_{3}$.

Finally, one gets by permutation symmetry the more compact expressions
\begin{align}
  \mathbb{E}_{n}\left[G_{1}^{2}\right]& =6\left(\frac{2\pi}{L}\right)^{2d}\sum_{\vec{\sigma}_{123}}\sum_{\vec{\bm{k}}_{123}}\left(\vec{\sigma}_{123}\cdot\vec{\lambda}_{123}\right)^{2}\left\lvert V_{\vec{\bm{k}}_{123}}^{\vec{\sigma}_{123}}\right\rvert^{2} \label{eq:symmetric_avg_G1_square} \\
  & \hspace{2cm} \times \left\lvert b_{\bm{k}_{1}}^{(0)}\right\rvert^{2}\left\lvert b_{\bm{k}_{2}}^{(0)}\right\rvert^{2}\left\lvert b_{\bm{k}_{3}}^{(0)}\right\rvert^{2}\left\lvert \Delta_{\Delta t}\left(\vec{\sigma}_{123}\cdot\vec{\omega}_{123}\right)\right\rvert^{2}\chi_{L}^{d}(\vec{\sigma}_{123}\cdot\vec{\bm{k}}_{123}) \nonumber
\end{align}
\begin{align}
  \mathbb{E}_{n}\left[G_{2}\right] & = 6 \left(\frac{2\pi}{L}\right)^{2d}\sum_{\vec{\sigma}_{123}}\sum_{\vec{\bm{k}}_{123}}\left\lvert V_{\vec{\bm{k}}_{123}}^{\vec{\sigma}_{123}}\right\rvert^{2}\chi_{L}^{d}(\vec{\sigma}_{123}\cdot\vec{\bm{k}}_{123})  \label{eq:symmetric_avg_G2} \\
                                   & \hspace{0.4cm} \times \left\{ \frac{1}{2}\left\lvert \Delta_{\Delta t}\left(\vec{\sigma}_{123}\cdot\vec{\omega}_{123}\right)\right\rvert^{2}
                                     \left( \sigma_{1}^{2}\lambda(\bm{k}_{1})\left\lvert b_{\bm{k}_{2}}^{(0)}\right\rvert^{2}\left\lvert b_{\bm{k}_{3}}^{(0)}\right\rvert^{2} \right. \right. \nonumber \\
                                   & \hspace{5cm}  +\sigma_{2}^{2}\lambda(\bm{k}_{2})\left\lvert b_{\bm{k}_{1}}^{(0)}\right\rvert^{2}\left\lvert b_{\bm{k}_{3}}^{(0)}\right\rvert^{2} \nonumber \\
                                   & \hspace{6cm}  \left.  +\sigma_{3}^{2}\lambda(\bm{k}_{3})\left\lvert b_{\bm{k}_{1}}^{(0)}\right\rvert^{2}\left\lvert b_{\bm{k}_{2}}^{(0)}\right\rvert^{2}\right) \nonumber \\
                                   & \hspace{1cm} +\tilde{E}_{\Delta t}\left(-\vec{\sigma}_{123}\cdot\vec{\omega}_{123};\vec{\sigma}_{123}\cdot\vec{\omega}_{123}\right) \nonumber \\
                                   & \hspace{2cm} \left. \times \left(\sigma_{1}\lambda(\bm{k}_{1})\left\lvert b_{\bm{k}_{1}}^{(0)}\right\rvert^{2}\left\lvert b_{\bm{k}_{2}}^{(0)}\right\rvert^{2}\sigma_{3}+\text{cyclic. perm.}\right)\right\} \nonumber
\end{align}

\paragraph{Expressions in terms of the empirical spectral density $n$}
Since the average is conditioned on $n(\bm{\xi})=\left(\tfrac{2\pi}{L}\right)^{d}\sum_{\bm{k}}\left\lvert b_{\bm{k}}^{(0)}\right\rvert^{2}\delta(\bm{\xi}-\bm{k})$,
one can replace all the discrete sums with respect to the wavevectors
$\bm{k}$ in (\ref{eq:symmetric_avg_G1_square}, \ref{eq:symmetric_avg_G2})
by continuous integrals:
\begin{align*}
\mathbb{E}_{n}\left[G_{2}\right] & =6\left(\frac{L}{2\pi}\right)^{d}\sum_{\vec{\sigma}_{123}}\int{\rm d}^{d}\xi_{1}\int{\rm d}^{d}\xi_{2}\int{\rm d}^{d}\xi_{3}\left\lvert V_{\vec{\bm{\xi}}_{123}}^{\vec{\sigma}_{123}}\right\rvert^{2}\chi_{L}^{d}\left(\vec{\sigma}_{123}\cdot\vec{\bm{\xi}}_{123}\right)\\
                                 & \hspace{0.3cm} \times\left\{ \tilde{E}_{\Delta t}\left(-\vec{\sigma}_{123}\cdot\vec{\omega}_{123};\vec{\sigma}_{123}\cdot\vec{\omega}_{123}\right)\left(\sigma_{1}\sigma_{3}\lambda(\bm{\xi}_{1})n(\bm{\xi}_{1})n(\bm{\xi}_{2})+\text{cyclic. perm.}\right)\right.\\
                                 & \hspace{0.8cm} \left.+\frac{1}{2}\left\lvert \Delta_{\Delta t}\left(\vec{\sigma}_{123}\cdot\vec{\omega}_{123}\right)\right\rvert^{2}\text{\ensuremath{\left(\sigma_{1}^{2}\lambda(\bm{\xi}_{1})n(\bm{\xi}_{2})n(\bm{\xi}_{3})+ \text{cyclic. perm.}
                                   \right)}}\right\} \\
\mathbb{E}_{n}\left[G_{1}^{2}\right] & =6\left(\frac{L}{2\pi}\right)^{d}\sum_{\vec{\sigma}_{123}}\int{\rm d}^{d}\xi_{1}\int{\rm d}^{d}\xi_{2}\int{\rm d}^{d}\xi_{3}\left\lvert V_{\vec{\bm{\xi}}_{123}}^{\vec{\sigma}_{123}}\right\rvert^{2} \chi_{L}^{d}\left(\vec{\sigma}_{123}\cdot\vec{\bm{\xi}}_{123}\right) \\
 & \qquad\times\left(\vec{\sigma}_{123}\cdot\vec{\lambda}_{123}\right)^{2}n(\bm{\xi}_{1})n(\bm{\xi}_{2})n(\bm{\xi}_{3})\left\lvert \Delta_{\Delta t}\left(\vec{\sigma}_{123}\cdot\vec{\omega}_{123}\right)\right\rvert^{2}\,.
\end{align*}

\paragraph{Asymptotic expressions of $\mathbb{E}_{n}\left[G_{1}^{2}\right]$,
$\mathbb{E}_{n}\left[G_{2}\right]$ and $Z_{L,\epsilon}$ in the kinetic
limit}
As explained in Sec. \ref{subsec:The-kinetic-limit}, the kinetic
limit refers to the joint limit $L\to\infty$ and $\epsilon\to0$
such that $L\epsilon^{2}$ is infinite or at least bounded from below
in order to get a large number of quasiresonances \cite{Nazarenko2011}.

Here, the time $\Delta t$ is chosen such that the inequality (\ref{eq:kinetic_ineq})
is fulfilled. Fixing $\Delta\tau=\mathcal{O}(1)$,
we have to evaluate $\mathbb{E}_{n}[G_{1}^{2}]$ and $\mathbb{E}_{n}[G_{2}]$
in the kinetic limit. The quantities $\mathbb{E}_{n}[G_{1}^{2}]$ and $\mathbb{E}_{n}[G_{2}]$
are of the kind
\[
\left(\frac{L}{2\pi}\right)^{d}\sum_{\vec{\sigma}_{123}}\int{\rm d}^{d}\xi_{1}\int{\rm d}^{d}\xi_{2}\int{\rm d}^{d}\xi_{3}X(\bm{\xi}_{1},\bm{\xi}_{2},\bm{\xi}_{3})\tilde{\chi}_{\Delta t}\left(\vec{\sigma}_{123}\cdot\vec{\omega}_{123}\right)\chi_{L}^{d}\left(\vec{\sigma}_{123}\cdot\vec{\bm{\xi}}_{123}\right)
\]
with $\tilde{\chi}_{\Delta t}(x)=\left\lvert \Delta_{\Delta t}(x)\right\rvert^{2}$ or $\tilde{\chi}_{\Delta t}(x)=\tilde{E}_{\Delta t}(-x,x)$, and $X(\bm{\xi}_1, \bm{\xi}_2 ,\bm{\xi}_3 )$
the remaining factors.

Proceeding formally, one can use the following asymptotic limits (see \cite[Eqs. (6.41, 6.42)]{Nazarenko2011})
\begin{align*}
\lim_{\Delta t\to\infty}\frac{\tilde{E}_{\Delta t}\left(-\vec{\sigma}\cdot\vec{\omega};\vec{\sigma}\cdot\vec{\omega}\right)}{\Delta t} & =\pi\delta\left(\vec{\sigma}\cdot\vec{\omega}\right)\\
\lim_{\Delta t\to\infty}\frac{\left\lvert \Delta_{\Delta t}\left(\vec{\sigma}\cdot\vec{\omega}\right)\right\rvert^{2}}{\Delta t} & =2\pi\delta\left(\vec{\sigma}\cdot\vec{\omega}\right)\\
\lim_{L\to\infty}\chi_{L}^{d}(\vec{\sigma}\cdot\vec{\bm{\xi}}) & =\delta\left(\vec{\sigma}\cdot\vec{\bm{\xi}}\right)
\end{align*}
to conclude that
\begin{align*}
  & \left(\frac{L}{2\pi}\right)^{d}\sum_{\vec{\sigma}_{123}}\int  \!\!  {\rm d}^{d}\xi_{1}  \!\!  \int  \!\!  {\rm d}^{d}\xi_{2}  \!\!  \int  \!\!  {\rm d}^{d}\xi_{3} X(\bm{\xi}_{1},\bm{\xi}_{2},\bm{\xi}_{3})\tilde{\chi}_{\Delta t}\left(\vec{\sigma}_{123}\cdot\vec{\omega}_{123}\right)\chi_{L}^{d}\left(\vec{\sigma}_{123}\cdot\vec{\bm{\xi}}_{123}\right) \\
  & \hspace{1cm} \underset{\text{kin.}}{\sim}c\pi\left(\frac{L}{2\pi}\right)^{d} \!\! \frac{\Delta\tau}{\epsilon^{2}}\sum_{\vec{\sigma}_{123}}\int  \!\!  {\rm d}^{d}\xi_{1}  \!\!  \int  \!\!  {\rm d}^{d}\xi_{2}  \!\!  \int  \!\!  {\rm d}^{d}\xi_{3} X(\bm{\xi}_{1},\bm{\xi}_{2},\bm{\xi}_{3})\delta(\vec{\sigma}_{123}\cdot\vec{\omega}_{123})\delta(\vec{\sigma}_{123}\cdot\vec{\bm{\xi}}_{123})
\end{align*}
with $c=1$ for $\tilde{\chi}_{\Delta t}(x)=\tilde{E}_{\Delta t}(-x,x)$
and $c=2$ for $\tilde{\chi}_{\Delta t}(x)=\left\lvert \Delta_{\Delta t}(x)\right\rvert^{2}$.
This can be seen with the choice $L\to\infty$ first, followed by
$\epsilon\to0$ as done for instance in \cite[Chap. 6]{Nazarenko2011}.

One finally obtains
\begin{align*}
  &\lim_{\Delta \tau \to 0}\text{Kin lim}\left\{ \left(\frac{2\pi}{L}\right)^{d}\frac{\epsilon^{2}}{\Delta\tau}\mathbb{E}_{n}\left[G_{2}\right]\right\}   \\
  & \hspace{1.5cm}  =6\pi\int{\rm d}^{d}\xi_{1}\lambda(\bm{\xi}_{1})\sum_{\vec{\sigma}}\int{\rm d}^{d}\xi_{2}\int{\rm d}^{d}\xi_{3}\left\lvert V_{\vec{\bm{\xi}}}^{\vec{\sigma}}\right\rvert^{2}\delta\left(\vec{\sigma}\cdot\vec{\omega}\right) \delta\left(\vec{\sigma}\cdot\vec{\bm{\xi}}\,\right)\\
 & \hspace{3cm} \times\left(\vec{\sigma}\cdot\vec{\lambda}\right)\left(\sigma_{1}n(\bm{\xi}_{2})n(\bm{\xi}_{3})+\sigma_{2}n(\bm{\xi}_{1})n(\bm{\xi}_{3})+\sigma_{3}n(\bm{\xi}_{1})n(\bm{\xi}_{2})\right) \\
  & \lim_{\Delta \tau \to 0} \text{Kin lim}\left\{ \left(\frac{2\pi}{L}\right)^{d}\frac{\epsilon^{2}}{\Delta\tau}\mathbb{E}_{n}\left[G_{1}^{2}\right]\right\} \\
  & \hspace{1.5cm}  =12\pi\sum_{\vec{\sigma}}\int{\rm d}^{d}\xi_{1}\int{\rm d}^{d}\xi_{2}\int{\rm d}^{d}\xi_{3}\left\lvert V_{\vec{\bm{\xi}}}^{\vec{\sigma}}\right\rvert^{2} \delta\left(\vec{\sigma}\cdot\vec{\omega}\right)\delta\left(\vec{\sigma}\cdot\vec{\bm{\xi}} \,\right)  \\
  & \hspace{4cm} \times \left(\vec{\sigma}\cdot\vec{\lambda}\right)^{2}  n(\bm{\xi}_{1})n(\bm{\xi}_{2})n(\bm{\xi}_{3})\;.
\end{align*}

Coming back to the expression of $Z_{L,\epsilon}$ (\ref{eq:mgf_Z_function_G1_G2}), the previous calculations allows
one to get
\begin{align*}
  &\lim_{\Delta \tau \to 0} \text{Kin lim}\left\{ \left(\frac{2\pi}{L}\right)^{d}\frac{\epsilon^{2}}{\Delta\tau}\left(\mathbb{E}_{n}\left[G_{2}\right]+\frac{1}{2}\mathbb{E}_{n}\left[G_{1}^{2}\right]\right)\right\}  \\
  & \hspace{1cm} = 6\pi\left\{ \sum_{\vec{\sigma}}\int{\rm d}^{d}\xi_{1}\int{\rm d}^{d}\xi_{2}\int{\rm d}^{d}\xi_{3}\left\lvert V_{\vec{\bm{\xi}}}^{\vec{\sigma}}\right\rvert^{2}\delta\left(\vec{\sigma}\cdot\vec{\omega}\right)\delta\left(\vec{\sigma}\cdot\vec{\bm{\xi}}\,\right)\right.\\
  & \hspace{1.5cm} \qquad\qquad +\left(\vec{\sigma}\cdot\vec{\lambda}\right)\left(\sigma_{1}n(\bm{\xi}_{2})n(\bm{\xi}_{3})+\sigma_{2}n(\bm{\xi}_{1})n(\bm{\xi}_{3})+\sigma_{3}n(\bm{\xi}_{1})n(\bm{\xi}_{2})\right)\\
  & \hspace{1.5cm} \qquad\qquad +\left.\left(\vec{\sigma}\cdot\vec{\lambda}\right)^{2}\times n(\bm{\xi}_{1})n(\bm{\xi}_{2})n(\bm{\xi}_{3})\right\} \;.
\end{align*}

Therefore, defining
\begin{align*}
  \mathcal{R}_{1}(L,\epsilon) & =\left(\frac{2\pi}{L}\right)^{d}\frac{\epsilon^{2}}{\Delta\tau}\left(\mathbb{E}_{n}\left[G_{2}\right]+\frac{1}{2}\mathbb{E}_{n}\left[G_{1}^{2}\right]\right) \\
  & \qquad -\lim_{\Delta \tau \to 0}\text{Kin lim}\left\{ \left(\frac{2\pi}{L}\right)^{d}\frac{\epsilon^{2}}{\Delta\tau}\left(\mathbb{E}_{n}\left[G_{2}\right]+\frac{1}{2}\mathbb{E}_{n}\left[G_{1}^{2}\right]\right)\right\} \, ,
\end{align*}
one finally obtains the asymptotic estimate (\ref{eq:perturbed_estimation_Z})
of $Z_{L,\epsilon}$:
\begin{align*}
Z_{L,\epsilon} & =1+\Delta\tau\left(\frac{L}{2\pi}\right)^{d}\left\{ 6\pi\sum_{\vec{\sigma}}\int{\rm d}^{d}\xi_{1}\int{\rm d}^{d}\xi_{2}\int{\rm d}^{d}\xi_{3}\left\lvert V_{\vec{\bm{\xi}}}^{\vec{\sigma}}\right\rvert^{2}\delta\left(\vec{\sigma}\cdot\vec{\omega}\right)\delta\left(\vec{\sigma}\cdot\vec{\bm{\xi}}\,\right)\right.\\
 & \qquad\qquad\qquad+\left(\vec{\sigma}\cdot\vec{\lambda}\right)\left(\sigma_{1}n(\bm{\xi}_{2})n(\bm{\xi}_{3})+\sigma_{2}n(\bm{\xi}_{1})n(\bm{\xi}_{3})+\sigma_{3}n(\bm{\xi}_{1})n(\bm{\xi}_{2})\right)\\
 & \qquad\qquad\qquad+\left.\left(\vec{\sigma}\cdot\vec{\lambda}\right)^{2} n(\bm{\xi}_{1})n(\bm{\xi}_{2})n(\bm{\xi}_{3})+\mathcal{R}_{1}(L,\epsilon)+\epsilon^{2}\mathcal{R}_{2}(L,\epsilon)\right\} \;.
\end{align*}

where $\mathcal{R}_{2}(L,\epsilon)$ corresponds to the contribution
of higher order ($\mathcal{O}(\epsilon^{4})$ and beyond) terms that
have not been considered in the perturbative expansion (\ref{eq:pert_exp_bk}).
As explained in Sec. \ref{subsec:derivation_large_deviations_Hamiltonian},
we will assume here that $\text{Kin lim }\epsilon^{2}\mathcal{R}_{2}(L,\epsilon)=0$.

\section{Quasipotential at equilibrium and entropy\label{sec:Appendix-Quasipotential_and_Entropy}}

In this appendix, we compute the equilibrium quasipotential for the
empirical spectral density. Its relation with the entropy associated
with the microcanonical measure (at fixed energy $E$ and momentum
$\bm{K}$) is discussed.

In order to regularize the ultraviolet divergence that occurs for
the Rayleigh-Jeans spectrum at equilibrium as well as to define properly
the microcanonical distribution, we assume that the set of wavenumbers
is restricted to $\mathbb{K}_{L}^{d}=\left\{ \bm{k}=\frac{2\pi}{L}\mathbb{Z}^{d}\left\lvert \left\vert  \bm{k}\right\vert  \leqslant k_{{\rm max}}\right.\right\} $.
Therefore, we consider only a finite number of modes $\mathcal{N}_{L}\sim\left(\tfrac{L}{2\pi}k_{{\rm max}}\right)^{d}$
in this appendix. For the sake of simplicity, we keep the notation
$\sum_{\bm{k}}$ to refer to $\sum_{\bm{k}\in\mathbb{K}_{L}^{d}}$.

We define the microcanonical distribution at fixed energy $E$ and
momentum $\bm{K}$ on the space of the re-scaled amplitudes $a_{\bm{k}}$:
\begin{equation}
\text{d}\mu_{E,\bm{K},L,\epsilon}=\frac{1}{\Gamma_{E,\bm{K},L,\epsilon}}\delta\left(E-\mathcal{\tilde{H}}\right)\delta\left(\bm{K}-\left(\tfrac{2\pi}{L}\right)^{d}\sum_{\bm{k}}\bm{k}\left\lvert a_{\bm{k}}\right\rvert^{2}\right)\prod_{\bm{k}}{\rm d}a_{\bm{k}}{\rm d}a_{\bm{k}}^{\ast},\label{eq:microcanonical_distribution}
\end{equation}
with $\Gamma_{E,\bm{K},L,\epsilon}(E,\bm{K})=\prod_{\bm{k}}\int{\rm d}a_{\bm{k}}{\rm d}a_{\bm{k}}^{\ast}\delta\left(E-\tilde{\mathcal{H}}\right)\delta\left(\bm{K}-\left(\tfrac{2\pi}{L}\right)^{d}\sum_{\bm{k}}\bm{k}\left\lvert a_{\bm{k}}\right\rvert^{2}\right)$
the volume of the phase space associated with the macrostate $(E,\bm{K})$.
We keep the indices $(L,\epsilon)$ in order to remember that the
energy $\tilde{\mathcal{H}}$ depends on $\epsilon$ and that the
system is of linear size $L$. The energy
$$
\mathcal{\tilde{H}}\equiv\epsilon^{-2}\mathcal{H}=\tilde{\mathcal{H}}_{2}+\epsilon\tilde{\mathcal{H}}_{3}
$$
is the microscopic energy $\mathcal{H}$ of the microscopic modes
(\ref{eq:H_dynamics}) rescaled by $\epsilon$. In terms of the rescaled
amplitudes $\left\lvert a_{\bm{k}}\right\rvert^{2}$ and the phases $\left\{ \varphi_{\bm{k}}\right\} ,$
we get
$$
\mathcal{\tilde{H}}_{2}=\left(\frac{2\pi}{L}\right)^{d}\sum_{\bm{k}}\omega_{\bm{k}} \vert a_{\bm{k}}\vert^{2}
$$
and
$$
\mathcal{\tilde{H}}_{3}=\left(\frac{2\pi}{L}\right)^{3d/2}\sum_{\vec{\sigma}}\sum_{\vec{\bm{k}}}V_{\vec{\bm{k}}}^{\vec{\sigma}}\left\lvert a_{\bm{k}_{1}}\right\rvert\left\lvert a_{\bm{k}_{2}}\right\rvert\left\lvert a_{\bm{k}_{3}}\right\rvert{\rm e}^{i\left(\varphi_{\bm{k}_{1}}+\varphi_{\bm{k}_{2}}+\varphi_{\bm{k}_{3}}\right)}\delta_{\vec{\sigma}\cdot\vec{\bm{k}},0}.
$$

Our goal is to estimate the probability distribution of the empirical
density $\hat{n}$,
that is a macroscopic state of the system, from the microcanonical
measure (\ref{eq:microcanonical_distribution}) with fixed energy $E$ and fixed momentum $\bm{K}$. This distribution
of the empirical spectral density $\hat{n}$ is denoted $P_{E,\bm{K},L,\epsilon}\left[n\right]=\mathbb{E}_{E,\bm{K}L,\epsilon}\left[\delta\left(\hat{n}-n\right)\right]$,
with $\mathbb{E}_{E,\bm{K},L,\epsilon}$ the average with respect
to the microcanonical distribution (\ref{eq:microcanonical_distribution}).
One obtains

\begin{equation}
P_{E,\bm{K},L,\epsilon}[n]=\frac{\Gamma_{E,\bm{K},L,\epsilon}[n]}{\Gamma_{E,\bm{K},L,\epsilon}}\,,\label{eq:microcanonical_proba_nL=00003Dn}
\end{equation}
where $\Gamma_{E,\bm{K},L,\epsilon}[n]$ is the volume of the
phase space associated with the macrostate $(E,\bm{K},n)$:
\[
\Gamma_{E,\bm{K},L,\epsilon}[n]=\prod_{\bm{k}}\int{\rm d}a_{\bm{k}}{\rm d}a_{\bm{k}}^{\ast}\delta\left(E-\tilde{\mathcal{H}}\right)\delta\left(\bm{K}-\left(\tfrac{2\pi}{L}\right)^{d}\sum_{\bm{k}}\bm{k}\left\lvert a_{\bm{k}}\right\rvert^{2}\right)\delta\left(n-\hat{n}\right).
\]

In the limit $L\to\infty$, it is natural to expect a large deviation
principle
\begin{equation}
Q_{E,\bm{K},\epsilon}[n]=-\lim_{L\to\infty}\left(\frac{2\pi}{L}\right)^{d}\log P_{E,\bm{K},L,\epsilon}[n]\,.\label{eq:definition_quasipotential_at_eq}
\end{equation}
Although $Q_{E,\bm{K},\epsilon}$ can be defined for any value of
$\epsilon$, we will compute $Q_{E,\bm{K}}=\lim_{\epsilon\to0}Q_{E,\bm{K},\epsilon}$
since it is the relevant contribution for the large deviations of
the empirical spectral density in the kinetic limit \eqref{eq:def_kinlim}.
At leading order ($\epsilon\to0$), the cubic correction to the
energy $\tilde{\mathcal{H}}=\tilde{\mathcal{H}}_{2}+\epsilon\tilde{\mathcal{H}}_{3}$
is vanishing and one gets
\begin{align*}
  \Gamma_{E,\bm{K},L}[n] & =\lim_{\epsilon\to0}\Gamma_{E,\bm{K},L,\epsilon} \\
  &=\prod_{\bm{k}}\int{\rm d}a_{\bm{k}}{\rm d}a_{\bm{k}}^{\ast}\delta\left(E-\tilde{\mathcal{H}}_{2}\right)\delta\left(\bm{K}-\left(\tfrac{2\pi}{L}\right)^{d}\sum_{\bm{k}}\bm{k}\left\lvert a_{\bm{k}}\right\rvert^{2}\right)\delta\left(n-\hat{n}\right).
\end{align*}

Using Eq. (\ref{eq:microcanonical_proba_nL=00003Dn}), the quasipotential is naturally expressed in terms
of the entropy $s_{E,\bm{K}}[n]=\lim_{L\to\infty}\left(\frac{2\pi}{L}\right)^{d}\log\Gamma_{E,\bm{K},L}[n]$
of the macrostate $(n,E,\bm{K})$ and the entropy $s_{E,\bm{K}}=\lim_{L\to\infty}\left(\frac{2\pi}{L}\right)^{d}\log\Gamma_{E,\bm{K},L}$
of the macrostate $(E,\bm{K})$ as
\[
Q_{E,\bm{K}}[n]=s_{E,\bm{K}}-s_{E,\bm{K}}[n].
\]
Since $Q_{E,\bm{K}}[n]\geqslant0$ and $\inf_{n}Q_{E,\bm{K}}=0$ by
definition, one obtains by contraction $s_{E,\bm{K}}=\max_{n}\left\{ s_{E,\bm{K}}[n]\right\} $.
The entropy $s_{E,\bm{K}}[n]$ can be computed by using the inverse
Laplace transform of the Dirac-$\delta$, or equivalently going to
the canonical ensemble. We thus considers the free energy
\begin{align}
f_{\beta,\bm{\mu}}\left[\lambda\right] & =-\lim_{L\to\infty}\left(\frac{2\pi}{L}\right)^{d}\log\left[\prod_{\bm{k}}\int{\rm d}a_{\bm{k}}{\rm d}a_{\bm{k}}^{\ast}{\rm e}^{-\left[\beta\sum_{\bm{k}}\omega_{\bm{k}}\left\lvert a_{\bm{k}}\right\rvert^{2}+\bm{\mu}\cdot\sum_{\bm{k}}\bm{k}\left\lvert a_{\bm{k}}\right\rvert^{2}+\sum_{\bm{k}}\lambda(\bm{k})\left\lvert  a_{\bm{k}}\right\rvert^{2}\right]}\right]\label{eq:free_energy}\\
 & =\int{\rm d}^{d}\xi\,\log\left(\frac{\beta\omega_{\bm{\xi}}+\bm{\mu}\cdot\bm{\xi}+\lambda(\bm{\xi})}{2\pi}\right)\;.\nonumber
\end{align}

The entropy $s_{E,\bm{K}}[n]$ is obtained as the Legendre-Fenchel
transform of the free energy \eqref{eq:free_energy} that is differentiable
everywhere on its domain.
One gets
\begin{align}
s_{E,\bm{K}}[n] & =\inf_{\beta,\bm{\mu},\lambda}\left\{ \beta E+\bm{\mu}\cdot\bm{K}+\int{\rm d}^{d}\xi\,\lambda(\bm{\xi})n(\bm{\xi})-f_{\beta,\bm{\mu}}\left[\lambda\right]\right\} \nonumber \\
 & =\begin{cases}
\int{\rm d}^{d}\xi\,\left[1+\log\left(2\pi\right)+\log n(\bm{\xi})\right] & \text{if }E=\int{\rm d}^{d}\xi\,\omega_{\bm{\xi}}n(\bm{\xi})\,,\;\bm{K}=\int{\rm d}^{d}\xi\,\bm{\xi}n(\bm{\xi})\\
-\infty & \text{otherwise}
\end{cases}.\label{eq:entropy_macrostate_EKn}
\end{align}
The constant term (bounded for $k_{\rm max}<\infty$) within $s_{E,\bm{K}}[n]$ can be
safely discarded because only difference of entropy matters. Looking
for the supremum (with respect to $n$) of the entropy (\ref{eq:entropy_macrostate_EKn}),
The entropy of the macrostate $(E,\bm{K})$ reads
\begin{equation}
s_{E,\bm{K}}=\int{\rm d}^{d}\xi\,\log n^{\ast}(\bm{\xi})\,,\label{eq:entropy_macrostate_EK}
\end{equation}
with
\[
n^{\ast}(\bm{\xi})=\left[\beta\omega_{\bm{\xi}}+\bm{\mu}\cdot\bm{\xi}\right]^{-1}
\]
the so called Rayleigh-Jeans spectrum \cite[Chap. 9]{Nazarenko2011},
and $(\beta,\bm{\mu})$ such that $E=\int{\rm d}^{d}\xi\:\omega_{\bm{\xi}}n^{\ast}(\bm{\xi})$,
$\bm{K}=\int{\rm d}^{d}\xi\,\bm{\xi}n^{\ast}(\bm{\xi})$.

This explicit expression of the equilibrium Rayleigh-Jeans spectrum
clearly shows the appearance of an \emph{ultraviolet catastrophe }\cite[Chap. 9]{Nazarenko2011}
that prevents the physical existence of such solution in absence of
any cut-off $k_{{\rm max}}$ on the wavenumbers. Indeed, considering
for instance the conserved quantity $\beta E+\bm{\mu}\cdot\bm{K}=\int {\rm d}^{d}\xi$,
one sees that the latter cannot remains finite if the set of allowed
wavenumbers is not of finite volume.

Finally, from the expressions of the entropies \eqref{eq:entropy_macrostate_EKn}
and \eqref{eq:entropy_macrostate_EK}, one obtains
\begin{equation}
Q_{E,\bm{K}}[n]=\begin{cases}
  -\int{\rm d}^{d}\xi\,\log\left(\frac{n(\bm{\xi})}{n^{\ast}(\bm{\xi})}\right) & \text{if }E=\int{\rm d}^{d}\xi\,\omega_{\bm{\xi}}n(\bm{\xi})\,,\;\bm{K}=\int{\rm d}^{d}\xi\,\bm{\xi}n(\bm{\xi})\\
+\infty & \text{otherwise}
\end{cases}\,.\label{eq:microcanonical_quasipotential_appendix}
\end{equation}
\red{Note that because $E=\int{\rm d}^{d}\xi\,\omega_{\bm{\xi}}n(\bm{\xi})$ and $\bm{K}=\int{\rm d}^{d}\xi\,\bm{\xi}n(\bm{\xi})$ are conserved and $\int n(\bm{\xi})/n^{\ast}(\bm{\xi}) \mathrm{d}^d\xi = \beta E + \bm{\mu}\cdot\bm{K}$ one can rewrite the quasipotential as
\begin{equation}
  Q_{E,\bm{K}}[n]=\begin{cases}
  \int{\rm d}^{d}\xi\,\left( \frac{n(\bm{\xi})}{n^{\ast}(\bm{\xi})} - 1 - \log\left(\frac{n(\bm{\xi})}{n^{\ast}(\bm{\xi})}\right) \right) & \text{if }E=\int{\rm d}^{d}\xi\,\omega_{\bm{\xi}}n(\bm{\xi})\,,\;\bm{K}=\int{\rm d}^{d}\xi\,\bm{\xi}n(\bm{\xi})\\
+\infty & \text{otherwise}
\end{cases}\, .
  \label{eq:microcanonical_quasipotential_appendix_2}
\end{equation}
The positivity and convexity of the quasipotential $Q_{E,\bm{K}}$ can then directly be deduced from the properties of the function $x\mapsto x-1 - \log(x)$ that has a single minimum at $x=1$.}

\end{appendices}

\bibliography{biblio_wave_turbulence}

\end{document}